\documentclass[%
reprint,
 amsmath,amssymb,
 aps,
pra,
notitlepage,
]{revtex4-2}
\usepackage{graphicx}
\usepackage{dcolumn}
\usepackage{bm}
\usepackage{hyperref}
\usepackage{textgreek}
\usepackage{xcolor}
\usepackage[caption=false]{subfig}

\begin{document}
\preprint{APS/123-QED}

\title{Switching fronts, plateaus and Kerr oscillations of counterpropagating light\\
in ring resonators}

\author{Graeme N. \surname{Campbell$^{1}$}}
\email{graeme.campbell.2019@uni.strath.ac.uk}
\author{Shuangyou \surname{Zhang$^{2}$}}
\author{Leonardo \surname{Del Bino$^{2,3}$}}
\author{Pascal \surname{Del'Haye$^{2,3}$}}
\author{Gian-Luca \surname{Oppo$^{1}$}}
\affiliation{$^1$SUPA and Department of Physics, University of Strathclyde, 
Glasgow, G4 0NG, Scotland, UK\\ 
$^2$Max Planck Institute for the Science of Light, 91058 Erlangen, Germany\\ 
$^3$Department of Physics, Friedrich Alexander University Erlangen-Nuremberg, 91058 Erlangen, Germany
}

\begin{abstract}
    We characterise stationary fronts and dark solitons for counterpropagating waves in micro-ring and fibre resonators with two input fields, normal dispersion and nonlocal coupling. These features are different from those in systems with local coupling in that their existence and stability are due to a careful balance of the areas of offset from homogeneous solutions. When scanning one of the two cavity detunings, stable solutions formed plateaus separated by two fronts are present in one of the counter-propagating fields with the power of the other field being homogeneous. Two front plateau solutions have a one-to-one correspondence to solutions of a Lugiato-Lefever equation at the unique Maxwell point. By defining effective detunings and for fixed values of the input powers where the fronts are found, we determine expressions for both the Maxwell point and the distance of the stable fronts as functions of detunings and input powers of both fields in good agreement with numerical simulations. For certain values of the detunings we find multi-stable states of plateaus with fronts, oscillating homogeneous states and non-oscillating homogeneous states of the counter-propagating fields. Robustness and parameter ranges of these unusual dynamical states coexisting with stable non-homogeneous front solutions are provided.
\end{abstract}


\maketitle


\section{Introduction}\label{sec:intro}
The physics of micro-ring resonators has gained significant interest over the last decade for their many applications such as octave spanning frequency combs \cite{PasquaziReview18} for use in telecommunication \cite{pfeifle2014coherent,pfeifle2015optimally} and spectroscopy \cite{suh2016microresonator,dutt2018chip}, as well as fundamental studies of dissipative pattern formation and temporal cavity solitons (TCS)\cite{grelu2015nonlinear}. The micro-ring resonator system is well described by the longitudinal version of the Lugiato-Lefever equation (LLE) \cite{lugiato1987spatial} in the form of a damped, driven nonlinear Schr\"odinger equation with cavity detuning. It originally described the transverse, dissipative spatial structures in passive optical systems with diffraction and was later adapted into a longitudinal form to describe temporal pattern formation along the cavity length \cite{haelterman1992dissipative,lugiato2018lugiato}. 

In this paper we study the interaction of two counterpropagating input fields in a normally dispersive micro-ring resonator, which is described by two nonlocally coupled equations of LLE form \cite{Kondratiev20,Skryabin:20}. The anomalous dispersion case has been investigated in \cite{Fan20} where the soliton blockade phenomenon was introduced. In Section II homogeneous steady state solutions of both fields are investigated and shown to undergo several bifurcations when the detunings are scanned. In Section III we characterize steady state solutions where one field has a homogeneous power while the other forms either a single dark TCS or power plateaus separated by sharp fronts. In Section IV we determine the parameter ranges of the existence and stability of these hybrid two-SF solutions, derive a semi-analytical expression of the distance of stationary SFs as a function of the cavity detunings and compare it successfully with numerical simulations. In particular we demonstrate that stationary solutions with two SF in one of the counterpropagating fields are strongly related to similar solutions in a single normally dispersive LLE at the Maxwell point. Such stationary states have been observed with single input laser setups where a counterpropagating field is induced by backscattering \cite{yu2021continuum,wang2021self}, where a connection with the Maxwell point is also made. Steep kinks connecting two stable homogeneous solutions in the presence of bistability have been studied extensively in diffusive systems where they are known as fronts \cite{Fife77}, in nonlinear optics of scalar fields where they are known as switching waves \cite{rosanov2002spatial,parra2016origin,garbin2017experimental}, and in systems with exchange symmetry where they are known as domain walls \cite{coullet1990breaking,oppo1999domain,oppo2001characterization,gilles2017polarization,garbin2021dissipative}. The system of interest here has exchange symmetry between the two counterpropagating fields. The hybrid solutions described in Section III display power plateaus separated by two kinks and do not reflect this exchange symmetry since one field is homogeneous and the other one is not. For this reason we prefer to label the kinks as `switching fronts' (SF) instead of `domain walls' which was preferred in for example \cite{wang2021self}. The solitonic (localized) aspect of these solutions is located in the SF and not of course in the power plateaus. For this reason we also avoid the use of the term `platicons' as being an unhelpful mixing of the localized aspect of solitons with the extended character of the homogeneous solutions. In Section V we derive a semi-analytical description of zero dispersion SF, and show that the zero dispersion SF solutions well approximate transient states with nonzero dispersion as they move towards stable two-SF states. In Section VI we show the presence of nonlinear oscillations of homogeneous states in a symmetry broken and nonlocal regime similar to those predicted in symmetric regimes \cite{woodley2018universal,woodleyPRL21}. We then identify a multi-stability of nonlinear oscillations with SF states and lowest power homogeneous stationary states. Conclusions, connection to experiments and applications are presented in Section VII.

\begin{figure}[hbtp]
    \centering
    {\includegraphics[width=0.45\textwidth]{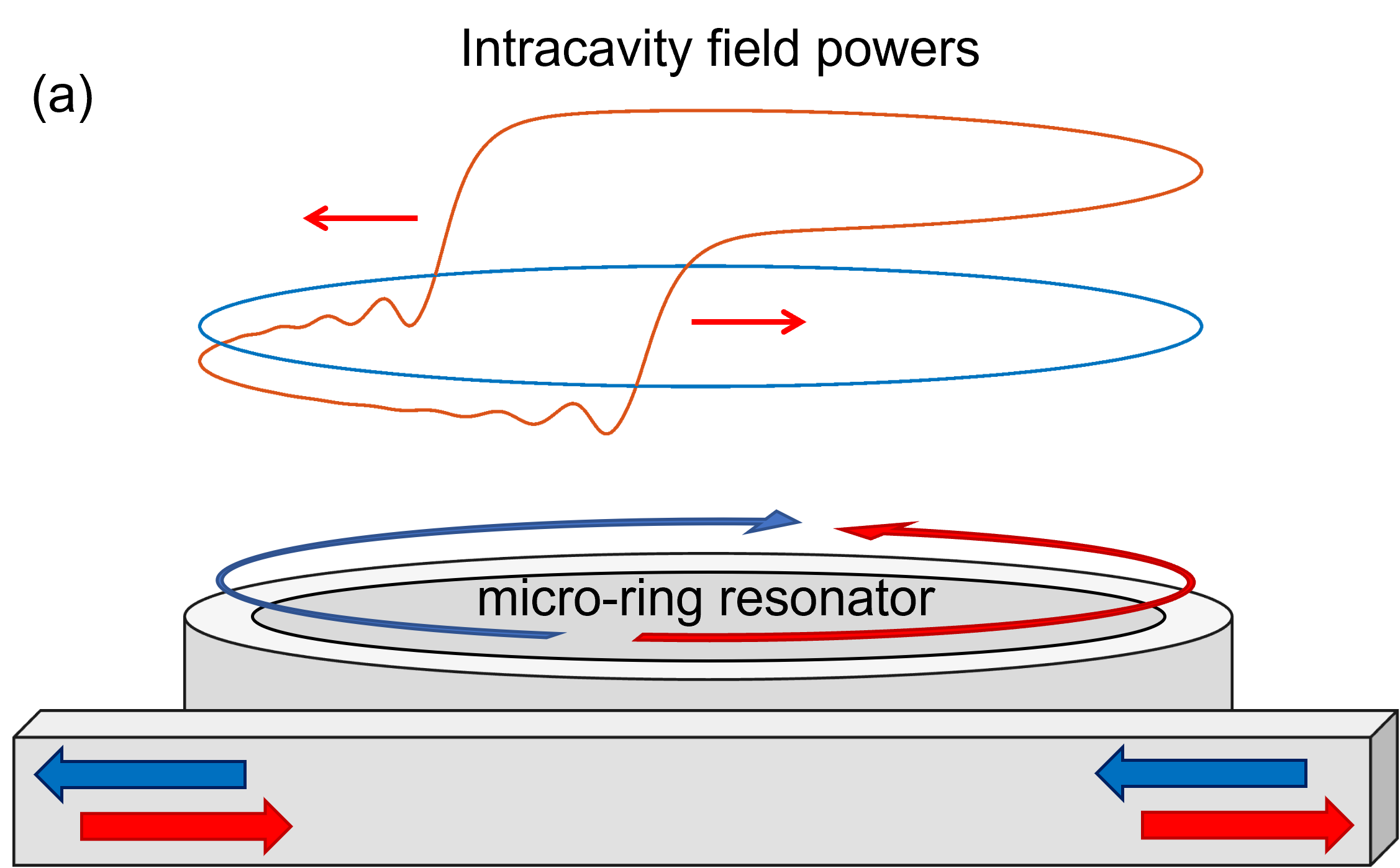}}
    {\includegraphics[width=0.25\textwidth]{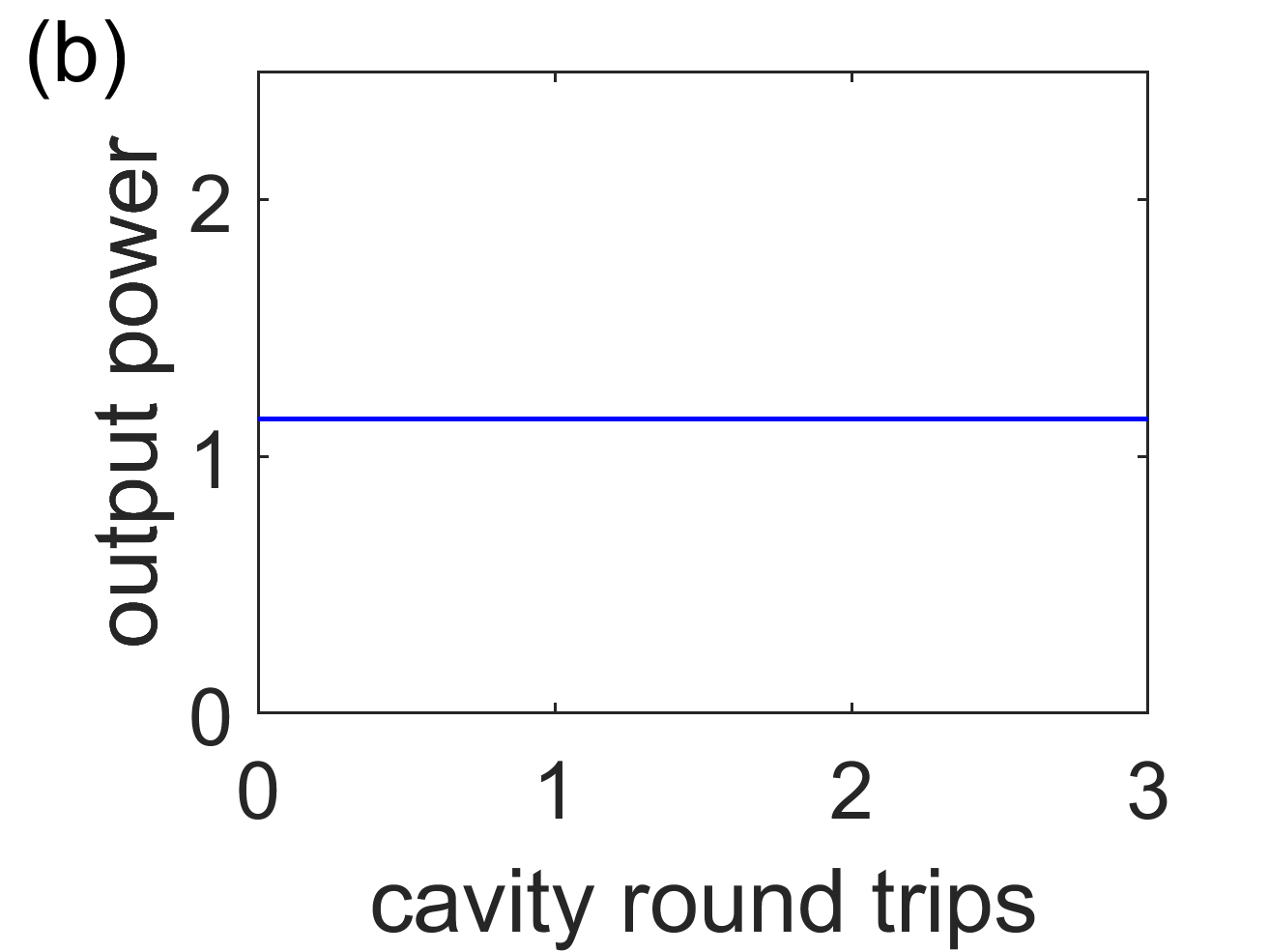}}{\includegraphics[width=0.25\textwidth]{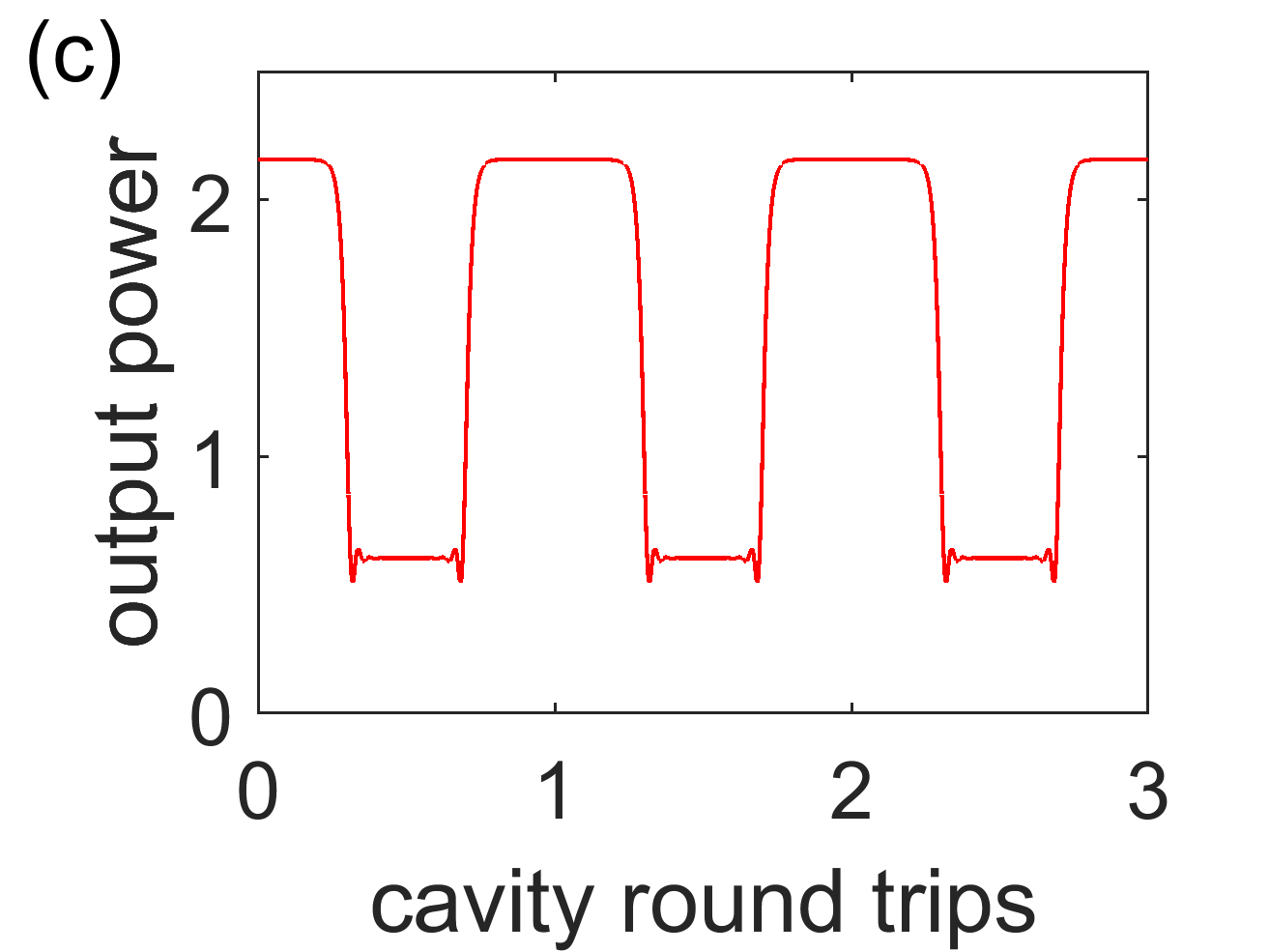}}
    \caption{A continuous wave (CW) forward (red) and a CW backward (blue) beams counterpropagate in a micro-ring resonator. For a detuning of the forward field smaller than the detuning of the backward field it is possible to obtain a power output where the backward field is still CW while the forward field displays two SFs in the intracavity power (a). This results in a switching output (c) from the forward field and CW output (b) from the backward field.}
    \label{fig:setup}
\end{figure}

\section{Nonlocally coupled counterpropagation in ring resonators}\label{sec:2LLE}
We consider the physical setting of a ring resonator pumped with two counterpropagating continuous wave (CW) lasers (see Fig. \ref{fig:setup}).
The two fields of this system are described by mean-field equations with the self- and cross-coupling terms in the Kerr approximation, through which the fields interact. Due to counterpropagation, the cross-coupling term is subject to the spatially averaged power of the counterpropagating field \cite{Kondratiev20,Skryabin:20}. The model for this system can be written in the adimensional, normalised form 
\begin{align}
    \partial_t F &= E_F - (1 + i\theta_F)F + i(|F|^2 + \nu\langle|B|^2\rangle)F - i\beta \partial_\zeta^2 F \label{eq:forwards} \\
    \partial_t B &= E_B - (1 + i\theta_B)B + i(|B|^2 + \nu\langle|F|^2\rangle)B - i \beta \partial_\zeta^2 B \label{eq:backwards}
\end{align}
where $t$ is the slow time over several round trips of the resonator, $F$ for forward and $B$ for backward are the complex amplitudes of the two counterpropagating fields in the ring resonator with identical polarisation, $E_F$ and $E_B$ are the input amplitudes, $\theta_F$ and $\theta_B$ the laser detunings from the nearest cavity resonance, $\nu$ the cross coupling coefficient that is in general equal to 2 for isotropic media, and $\zeta$ is the fast time variable over the round-trip. The last term describes normal dispersion with a positive dispersion coefficient $\beta$ while the power averages $\langle|F|^2\rangle$ and $\langle|B|^2\rangle$ are given by
\begin{align}
    \langle |F|^2 \rangle &= \frac{1}{L}\int_0^L|F|^2d\zeta\\
    \langle |B|^2 \rangle &= \frac{1}{L}\int_0^L|B|^2d\zeta
    \label{eq:integrals}
\end{align}
where $L$ is the length of the resonator. The configuration and parameters used here differ from those used in \cite{yu2021continuum,wang2021self} in that we consider energy injection on both fields. It is important to note that for $E_F=E_B$ and $\theta_F=\theta_B$ the system is perfectly symmetric upon the exchange of the forward and backward fields.  

\subsection{Homogeneous steady states}\label{subsec:HSS}
The homogeneous steady-state solutions (HSS) of counterpropagating fields are identical to the two polarization co-propagating regimes seen in \cite{woodley2018universal,hill2020effects} due to the cross terms containing $\langle|F|^2\rangle = |F|^2$, $\langle|B|^2\rangle = |B|^2$. Eqs. (\ref{eq:forwards})-(\ref{eq:backwards}) can be expressed by the coupled cubic equations
\begin{align}
    P_F &= H_F^3 -2(\theta_F - \nu H_B) H_F^2 + ((\theta_F - \nu H_B)^2 + 1)H_F \label{eq:HSSF}\\
    P_B &= H_B^3 -2(\theta_B - \nu H_F) H_B^2 + ((\theta_B - \nu H_F)^2 + 1)H_B \label{eq:HSSB}
\end{align}
where $H_F = |F|^2$, $H_B = |B|^2$ (the letter $H$ referring to the power of the HSS) while $P_F = |E_F|^2$ and $P_B = |E_B|^2$ correspond to the input powers. 

\begin{figure}[hbtp]
    \centering
    {\includegraphics[width=0.5\textwidth]{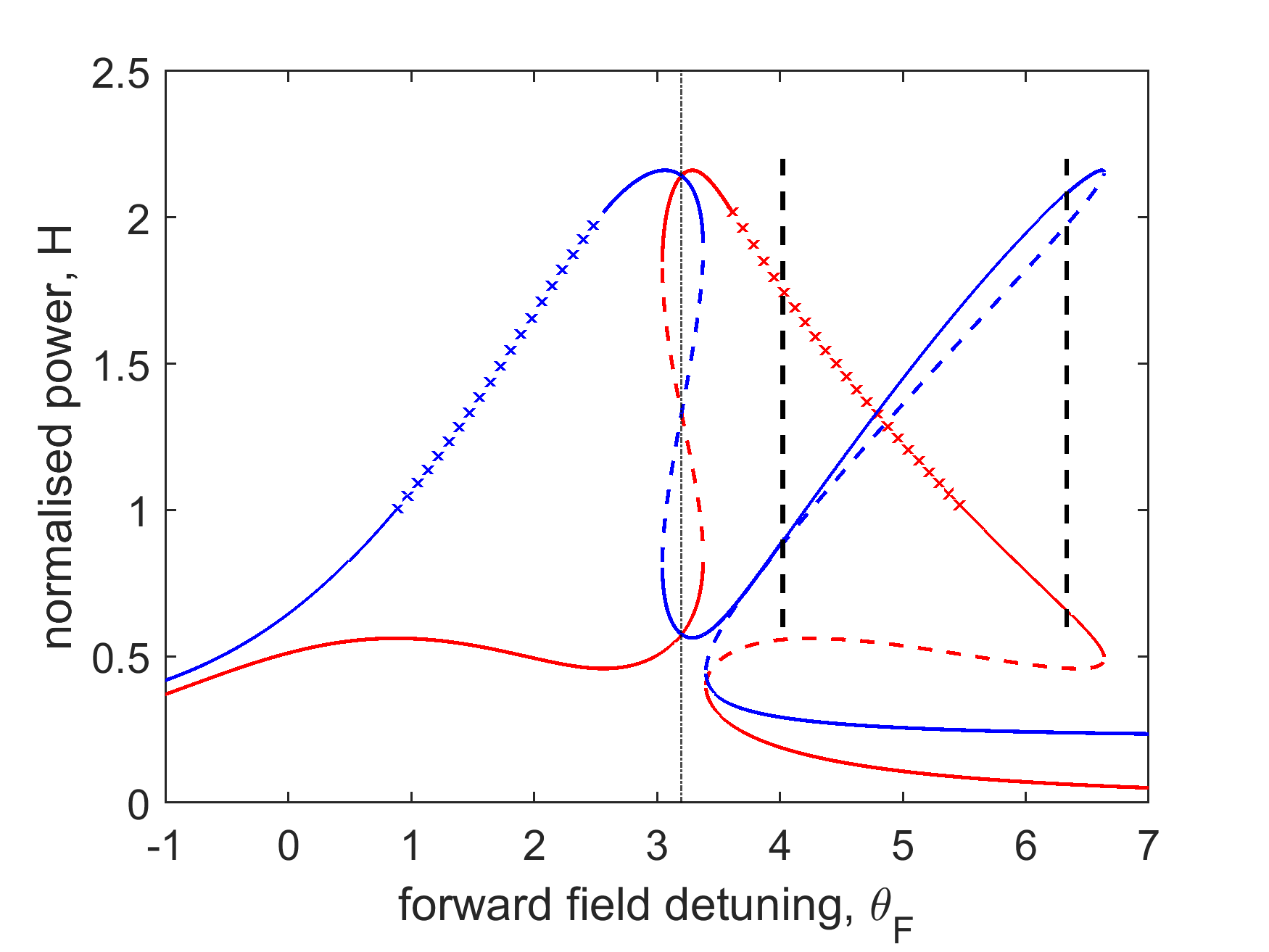}}
    \caption{
    Powers $H_F$ (red) and $H_B$ (blue) of HSS (\ref{eq:HSSF},\ref{eq:HSSB}) when changing the detuning $\theta_F$ of the forward field for parameter values $E_F=E_B=1.47$, $\nu=2,$ and the detuning of the backwards field kept constant at $\theta_B=3.2$. The solid (dashed) lines correspond to stable (unstable) HSS, the lines marked with the symbol X correspond to HSS unstable to fast time perturbations, the vertical black dashed lines correspond to Hopf bifurcations of the HSS.}
    \label{fig:HSS}
\end{figure}

These algebraic equations can be solved numerically for given values of the parameters, an example of which is shown in Fig. \ref{fig:HSS} for $\nu=2$, equal pump powers ($P_F=P_B=2.1609$) with one of the field detuning kept constant ($\theta_B = 3.2$) while the other ($\theta_F$) is changed. In the vicinity of equal detunings (dashed line) where the equations are symmetric upon exchange of the forward and backward fields, a bistability regime with a `figure 8' shape exists. Here we expect the `middle' HSS to be unstable (see dashed lines in Fig. \ref{fig:HSS}). When increasing the forward detuning $\theta_F$ after the symmetric value $3.2$, the figure of 8 ends in this case at the point where two new HSS are born in a degenerate saddle-node bifurcation, the lowermost being stable and the intermediate unstable. For values of $\theta_B<0.32$ the saddle-node bifurcation takes place after the end of the figure of 8, while for values of $\theta_B>0.32$ the saddle-node bifurcation takes place before the end of the figure of 8 leading to a simultaneous presence of 5 different stationary states. After the saddle-node bifurcation and the end of the figure of 8, multi-stability of homogeneous solutions is present at large values of the detuning $\theta_F$ until a reverse saddle-node bifurcation restores a single HSS at very large values of the scanned detuning. 

In the asymmetric region for $\theta_F>\theta_B$ we detect Hopf bifurcations of the HSS leading to oscillations as described in Section \ref{sec:Oscill}. The two Hopf bifurcations occur on the upper branches of the HSS (see the vertical dashed lines in Fig. \ref{fig:HSS}) and have opposite directions when increasing the detuning $\theta_F$, with the amplitude of the oscillation growing from around $\theta_F=4$ and decreasing to zero around $\theta_F=6.3$. These forward and backward Hopf bifurcations are analogous in nature and stability eigenvalues to those described in \cite{woodley2018universal,hill2020effects} where, however, the two detunings where kept equal to each other during the scan to focus on symmetric HSS.

There are however further instabilities of the HSS due to the nonlocal nature of Eqs. (\ref{eq:forwards})-(\ref{eq:backwards}). In Appendix A, a linear stability analysis of the HSS to inhomogeneous perturbations at zero dispersion on the fast time scale is presented. A new set of stability eigenvalues is found:
\begin{eqnarray}
    \lambda &=& -1 \pm \sqrt{-A_1 B_1}\\
    \lambda &=& -1 \pm \sqrt{-A_2 B_2}
    \label{NewStabEigen}
\end{eqnarray}
where $A_1 = H_F + \nu H_B - \theta_F$, $A_2 = H_B + \nu H_F - \theta_B$, $B_1 = 3H_F + \nu H_B - \theta_F$, $B_2 = 3H_B + \nu H_F - \theta_B$, with $H_F$ and $H_B$ being obtained from Eqs. (\ref{eq:HSSF})-(\ref{eq:HSSB}). These new eigenvalues are entirely due to the non-local terms of our system which means that local perturbations result in changes to the unperturbed regions. The lines marked with the letter X in Fig. \ref{fig:HSS} correspond to the HSS instabilities to inhomogeneous perturbations where the real part of one of the four eigenvalues (\ref{NewStabEigen}) is positive. Here the system, in general, evolves to two-SF steady solutions as described in Section III. Note however that HSS instabilities to inhomogeneous perturbations can affect regions where the HSS are also unstable to oscillations (see the right hand side of Fig. \ref{fig:HSS}).

\section{Two switching fronts and dark soliton steady states}
\begin{figure}[]
    \centering
    {\includegraphics[width=0.5\textwidth]{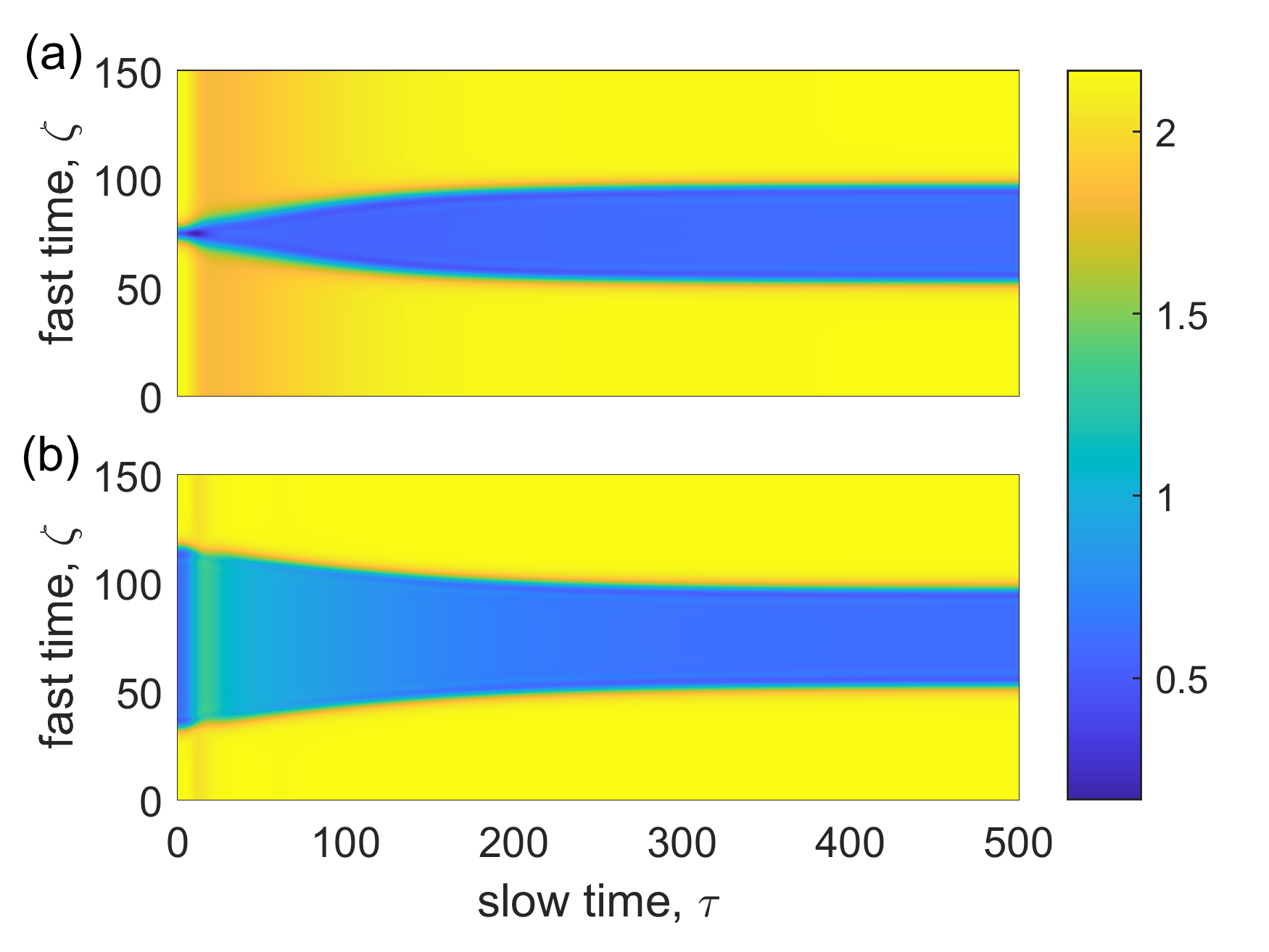}}
    \caption{Temporal evolution of the backward power towards a stable SF state for $|E_F|^2=|E_B|^2=2.1609$, $\theta_F=2.0,\theta_B=3.2$ from two different initial conditions with dispersion $\beta=1$. (a) Initial condition with two kinks at narrow separation. (b) Initial condition with two kinks at wide separation.}
    \label{fig:movingfronts}
\end{figure}

In the counterpropagating system with nonlocal coupling described by Eq. (\ref{eq:forwards})-(\ref{eq:backwards}), we observe the formation of steady states made of power plateaus separated by SF in one of the two counterpropagating fields while the second field remains homogeneous, for wide ranges of the detuning values. In Fig. \ref{fig:movingfronts} we show the formation of stable SF states when starting from a narrow (a) or broad (b) perturbation of the HSS for $E_F=E_B=1.47$, $\nu=2$, $\theta_B=3.2$ and $\theta_F=2.0$. In Section \ref{subsec:HSS} we showed that in this parameter region, HSS are unstable to inhomogeneous perturbations. In both cases of broad and narrow initial perturbations, the system evolves to the same final state formed by a SF state with a well-defined separation of the two SFs. It is important to note that the SF solutions do not connect HSS of the Eq. (\ref{eq:forwards})-(\ref{eq:backwards}) and affect only one of the counterpropagating fields, the other being homogeneous. They do not correspond to symmetry exchanges of the $F$ and $B$ fields.   

\begin{figure}[]
    \centering
    \includegraphics[width=0.25\textwidth]{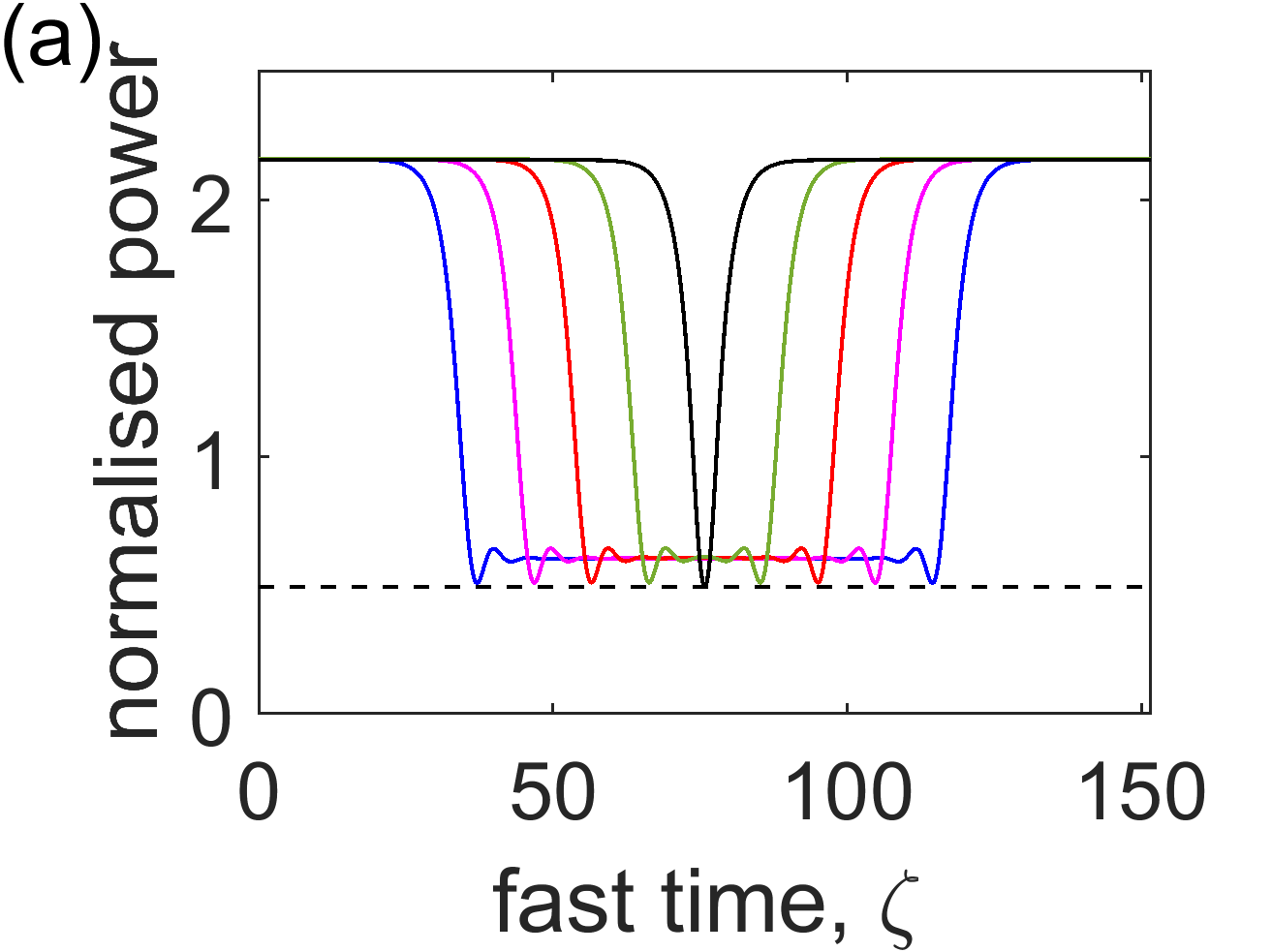}\includegraphics[width=0.25\textwidth]{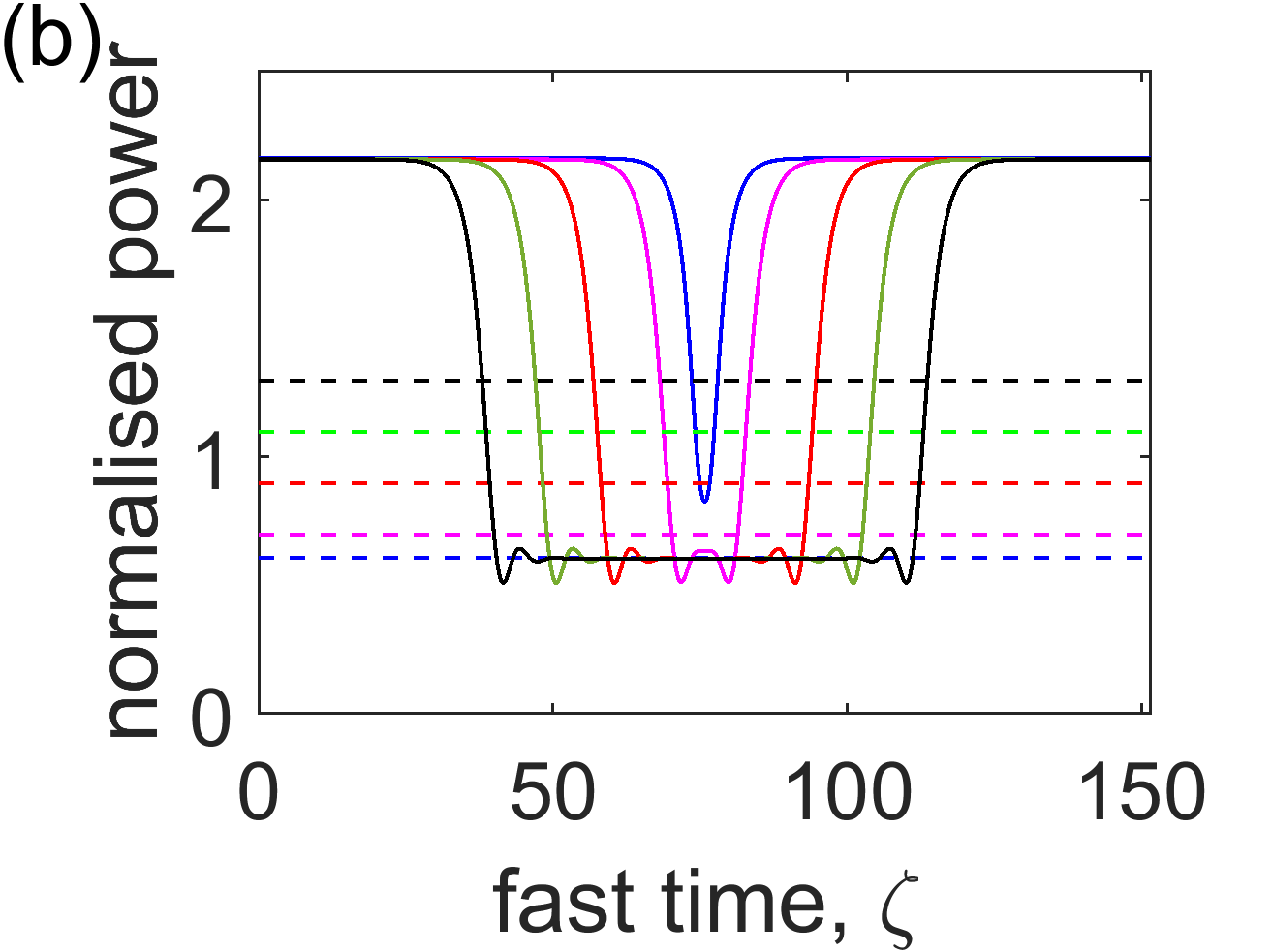}
    \caption{Various SF states for $E_F=E_B=1.47$, $\nu = 2$, $\beta=1$, $\theta_B = 3.2$. (a) Backward (forward) field power of steady state solutions, solid lines (dashed lines), for five values of $\theta_F = 1.2$ (blue), $\theta_F = 1.6$ (magenta), $\theta_F = 2.0$ (red), $\theta_F = 2.4$ (green), $\theta_F = 2.8$ (black). (b) Forward (backward) field power of steady state solutions, solid lines (dashed lines) for five values of $\theta_F = 3.4$ (blue), $\theta_F = 3.6$ (magenta), $\theta_F = 4.0$ (red), $\theta_F = 4.4$ (green), $\theta_F = 4.8$ (black).}
    \label{fig:difftheta}
\end{figure}

A number of stable asymptotic states are presented in Fig. \ref{fig:difftheta} for the same values of the parameters as Fig. \ref{fig:movingfronts} but with $\theta_F$ varying from 1.2 to 4.8. In the interval $1.2<\theta_F<2.8$ the backward (forward) intracavity power is non-homogeneous (homogeneous), Fig. \ref{fig:difftheta}a, while in the interval  $3.4<\theta_F<4.8$ the forward (backward) intracavity power is non-homogeneous (homogeneous), Fig. \ref{fig:difftheta}b. The solid lines correspond to the power profiles of the field where a dark structure is found while the dashed lines correspond to fully homogeneous solutions. When the two detunings are close to each other (for example $\theta_F$ equal to 2.8 or 3.4 in Fig. \ref{fig:difftheta}) the inhomogeneous field has the shape of a localized dark soliton. In the interval of $2.8<\theta_F<3.4$, there are no inhomogeneous stable solutions and the system relaxes to the HSS seen in Fig. \ref{fig:HSS}. This instability of the dark soliton solution is affected by the dispersion of the field and dark solitons can persist in larger detuning ranges for $\beta<1$. For the present choice of parameter values there is no bistability between the two SF states close to detuning symmetry. We will see in Section IV that for $P_F=P_B=3$, for example, an overlap region where both SF states are stable, exists. In this overlap region, bistability of SF states is observed where SFs are present in either the forward or the backward field with the other field homogeneous for the same parameter values.

When the two detunings are very different from each other, the inhomogeneous field can take the shape of a localised bright soliton while the other field remains homogeneous. Such states have been observed in single laser setups \cite{yu2021continuum}.

Stable SF states and stable dark solitons are present due to the nonlocal coupling of the two counterpropagating fields. Nonlocality of the cross coupling results in a shift in detuning of the fields. To this end we define effective detunings
\begin{eqnarray}
    \Tilde{\theta}_F &=& \theta_F - \nu\langle |B|^2\rangle \label{eq:effF}\\
    \Tilde{\theta}_B &=~& \theta_B - \nu\langle |F|^2\rangle \label{eq:effB}
\end{eqnarray}
that reduce the counterpropagating Eq. (\ref{eq:forwards})-(\ref{eq:backwards}) to a pair of LLEs nonlocally coupled via their effective detuning:
\begin{eqnarray}
    \partial_t F &= E_F - (1 + i\Tilde{\theta}_F)F + i|F|^2F - i\beta \partial_\zeta^2 F\\
    \partial_t B &= E_B - (1 + i\Tilde{\theta}_B)B + i|B|^2B - i \beta \partial_\zeta^2 B \, .
\end{eqnarray}
Taken separately when ignoring the coupling through the effective detunings, each of these LLEs displays a Maxwell point for normal dispersion corresponding to a set of parameter values where solutions made of power plateaus well separated by SFs are stable. For any other parameter value close to the Maxwell point, SFs are observed to move close or away from each other. At the Maxwell point and at the Maxwell point only, the LLE displays a multi-stability of power plateaus solutions with two stationary SFs at arbitrary separations. In gradient systems the Maxwell point corresponds to the parameter value where both bistable homogeneous states have equal energy. In non-gradient system, such as the LLE, Maxwell points and hysteresis can still be possible even though an expression of the energy cannot be obtained.

There are very important differences between our SF states and dark solitons due to nonlocal coupling and structures of similar shape in the single LLE with normal dispersion (at the Maxwell point or close to the Maxwell point) studied theoretically in \cite{parra2016origin,parra2016dark} and experimentally in \cite{xue2015mode,garbin2017experimental,XueWeiner2017}. For example, the power of the homogeneous field and the power values of the plateaus before and after the two SFs in the inhomogeneous field are not the values of the HSS studied in Section \ref{subsec:HSS}. When the values of the two field detunings are well separated, stable SF states are not due to locking mechanisms of the tails of the SFs as for example observed in optical parametric oscillators \cite{oppo1999domain,oppo2001characterization}. However, when the detunings of the two fields are quite close to each other, dark solitons owe their stability to the local oscillations in the lower part of the SF as shown in Fig. \ref{fig:difftheta} for $\theta_B=3.2$ and $\theta_F=2.8$, $3.4$ and $3.6$.

When increasing the detuning $\theta_F$ while keeping the detuning $\theta_B$ fixed, one observes first a decreasing separation between the two stable SFs in the backward field, Fig. \ref{fig:difftheta}a and then, after the symmetric state $\theta_F=\theta_B$, an increasing distance between the two stable SFs in the forward field as seen in Fig. \ref{fig:difftheta}b. In the latter case, the power of the homogeneous backward field changes substantially upon variations of $\theta_F>\theta_B$ while the power of the homogeneous forward field changes only a little upon variations of $\theta_F<\theta_B$ (see Fig. \ref{fig:difftheta}a). This effect is a direct result of the effective detunings that contain the integrals (\ref{eq:integrals}).

One very interesting feature when scanning one of the detunings (say $\theta_F$) while keeping the other one fixed by changing the input frequency of one of the two pumps, is that upon crossing the symmetric state $\theta_F=\theta_B$, stable SFs and dark solitons switch from one propagation direction (the backward for $\theta_F<\theta_B$) to the other (the forward for $\theta_F>\theta_B$). This provides the operator of this device to select at will the direction, in which the solitary structures and, consequently, an optical frequency comb occurs. 


\section{Distance of two stationary switching fronts}\label{sec:TFSS_distance}

From numerical simulations we obtain stationary solutions with two stable SFs separated by a distance $\Delta$. We aim here to obtain an analytical expression of the the distance $\Delta$ when using $\theta_F$ as a control parameter.

\begin{figure}[]
    \centering
    \includegraphics[width=0.5\textwidth]{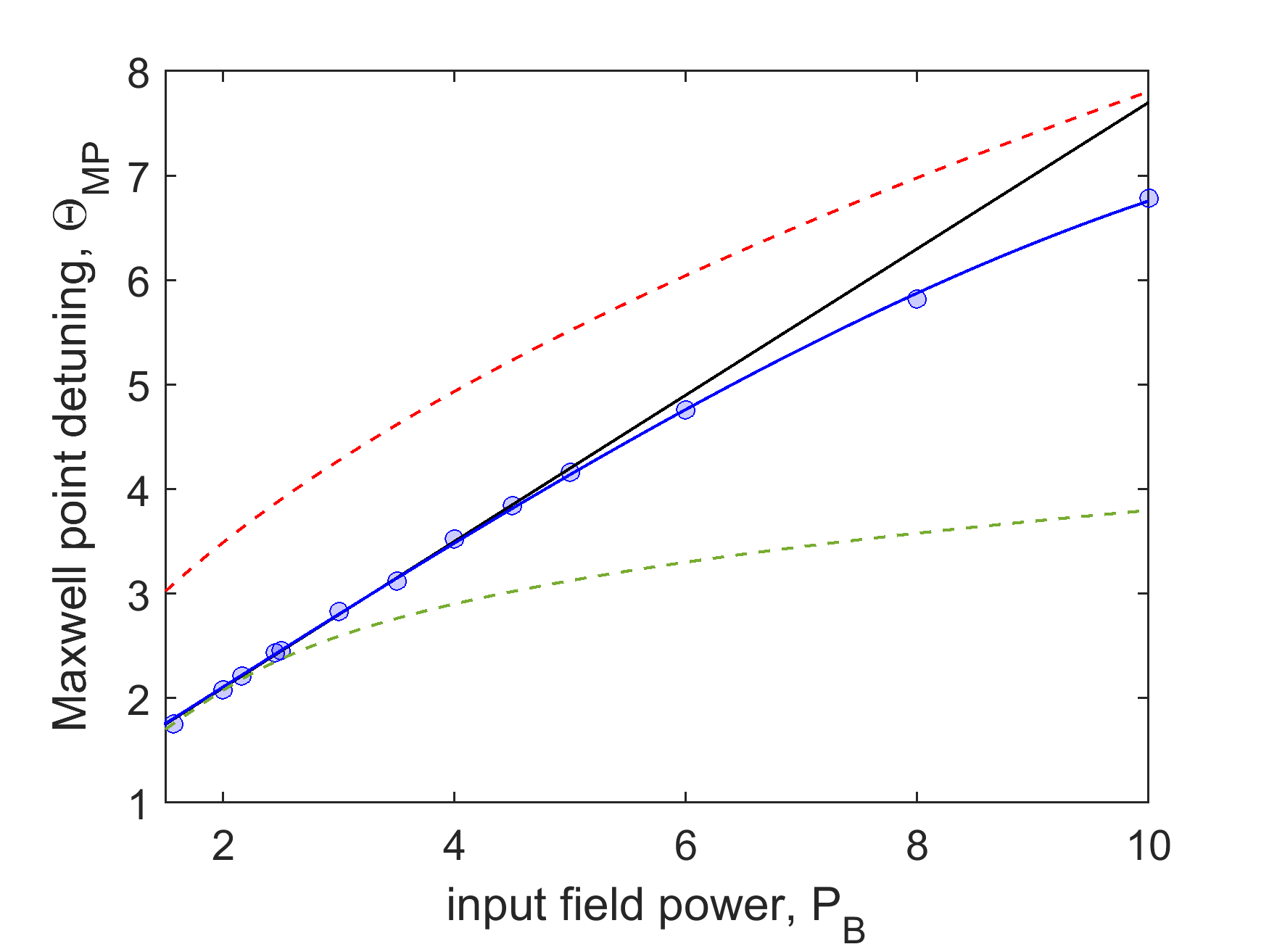}
    \caption{The detuning $\Theta_{MP}$ of a single LLE (\ref{eq:SSLL}) at the Maxwell point as a function of $P_B=E_B^2$. The circles are numerically evaluated points from which we obtain the linear (Eq. \ref{eq:TMP} in black) and cubic (Eq. \ref{eq:TMP3order} in blue) fitted curves for the Maxwell point distribution. The dashed green and dashed red curves correspond to the asymptotic and variational methods of \cite{wang2021self}, respectively.}
    \label{fig:MP}
\end{figure}

We start from the case of two SFs in the backward field for a given value of $\theta_B$ when changing $\theta_F<\theta_B$ (see Fig. \ref{fig:difftheta}a). In this case the forward field power $|F|^2$ is homogeneous and appears to be independent of the detuning $\theta_F$. Note that this homogeneous value of the forward power is not the HSS value $H_F$ discussed in Section III. For the stationary solutions we can write:
\begin{eqnarray}
    E_F &=& (1 + i\Tilde{\theta}_F) F - i|F|^2 F \label{eq:SSforwards}\\
    E_B &=& (1 + i\Tilde{\theta}_B) B - i|B|^2 B + i \beta \partial_\zeta^2 B \label{eq:SSbackwards}
\end{eqnarray}
where we have used Eqs. (\ref{eq:effF}) and (\ref{eq:effB}).
Each solution of the backward field equation (\ref{eq:SSbackwards}) when changing $\theta_F$ has a one to one correspondence with one of the multi-stable stationary solutions of a single Lugiato-Lefever equation (LLE) at the Maxwell point given by 
\begin{equation}
  E_B = (1 + i\Theta_{MP}) B - i|B|^2 B + i \beta \partial_\zeta^2 B 
  \label{eq:SSLL}
\end{equation}
where $\Theta_{MP}$ is the cavity detuning at Maxwell point which depends on the input power $P_B$. The functional dependence of $\Theta_{MP}$ from $P_B$ can be obtained by asymptotic methods close to the critical detuning value $\sqrt{3}$ for $P_B \approx 8\sqrt{3}/9$ and by variational methods for $P_B>10$ \cite{wang2021self}. Neither of these approximations is satisfactory in the range $2<P_B<7$ of values used here (see Fig.~\ref{fig:MP}). 
By computing the Maxwell points numerically (see blue line in Fig.~\ref{fig:MP}) we find that a simple linear dependence of $\Theta_{MP}$ from $P_B$
\begin{equation}
    \Theta_{MP} \approx \eta (1+P_B) 
    \label{eq:TMP}
\end{equation}
with $\eta=0.7$ approximates the numerical values much better in the interval of interest (see black line in Fig. \ref{fig:MP}). Additional terms can be included in the approximation to extend the range of validity to $P_B=10$
\begin{equation}
    \Theta_{MP} \approx \eta(-0.001997P_B^3 +0.006503P_B^2 + P_B +1)
    \label{eq:TMP3order}
\end{equation}

\begin{figure}[]
    \centering
    \includegraphics[width=0.5\textwidth]{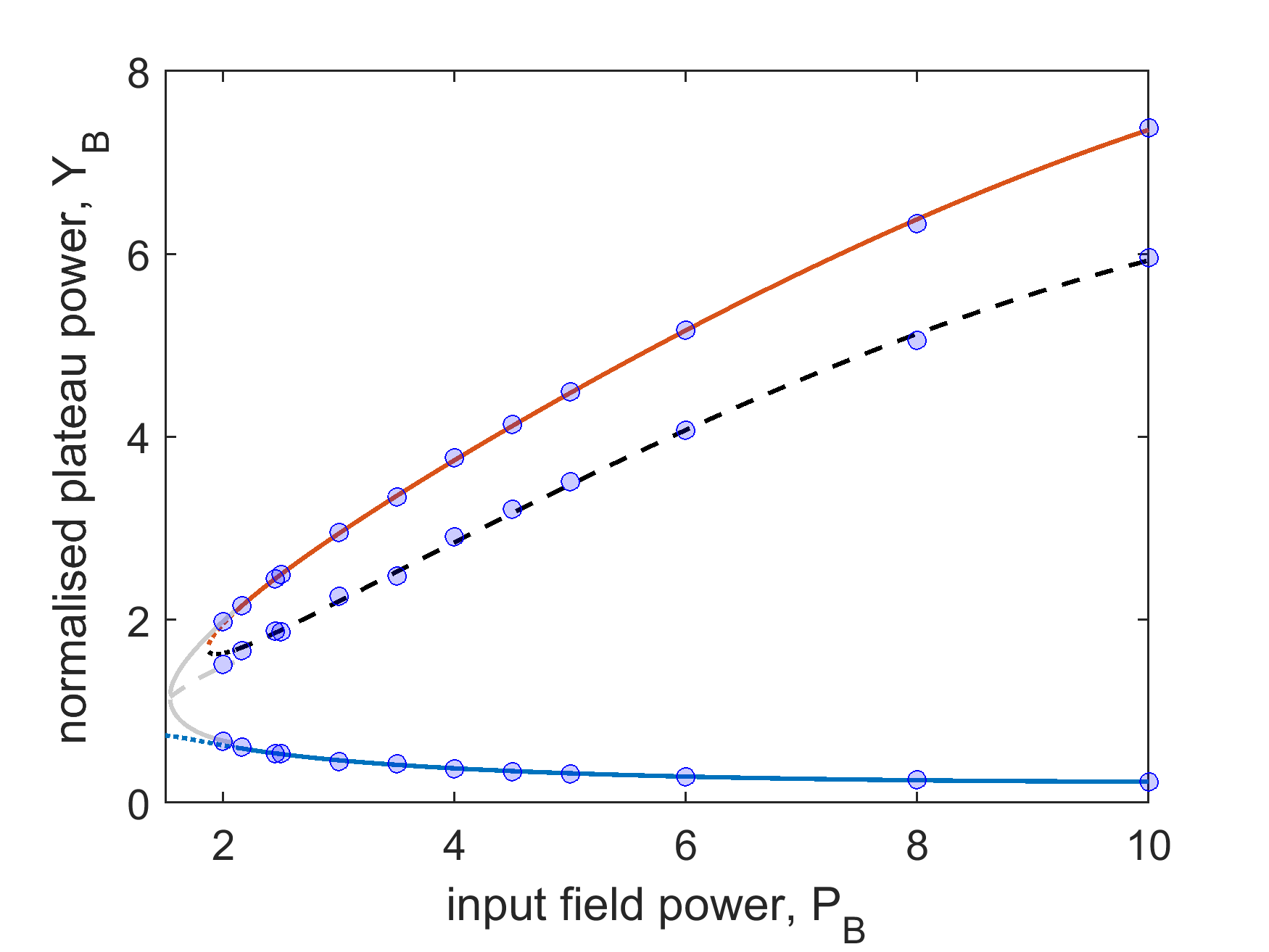}
    \caption{Power $Y_B$ of the homogeneous solutions before and after a SF for a single LLE at Maxwell point (\ref{eq:Y_B}). SFs are possible after the onset of bistability at critical pump power $P_B\approx 8\sqrt{3}/9$ where the higher $Y_B^+$ (red) and the lower $Y_B^-$ (blue) branches are stable but the dashed middle solution is unstable. 
    The Maxwell point detunings $\Theta_{MP}$ are approximated by (\ref{eq:TMP3order}) for $P>2.1$ and by an asymptotic approach for $P<2.1$. The blue circles are homogeneous solutions before and after a SF from the simulation of (\ref{eq:forwards})-(\ref{eq:backwards}).}
    \label{fig:tails}
\end{figure}

By using the equivalence between (\ref{eq:SSbackwards}) and (\ref{eq:SSLL}) as well as the definition of $\Tilde{\theta}_B$ in (\ref{eq:effB}) we obtain the value of the power of the homogeneous forward field for the SF state in the backward field:
\begin{align}
    \langle |F|^2\rangle = |F|^2 
    = \frac{1}{\nu}[\theta_B -  \Theta_{MP}] \approx \frac{1}{\nu}[\theta_B - \eta (1+P_B)]
    \label{eq:Fsquare}
\end{align}
As shown in the numerical simulations of the two SFs for $\theta_F<\theta_B=3.2$ in Fig. \ref{fig:difftheta}(a), $|F|^2$ is independent of the control parameter $\theta_F$ and its value is just below 0.5 for the case of $P_B=2.1609$, in agreement with (\ref{eq:Fsquare}). The power $Y_B=|B|^2$ of the homogeneous states of (\ref{eq:SSLL}) satisfies
\begin{align}
Y_B^3-2 \Theta_{MP} Y_B^2 + \left( 1+\Theta_{MP}^2 \right) Y_B - P_B = 0
\label{eq:Y_B}
\end{align} 
from which it is possible to obtain the values of the plateau powers $Y_B^+$ and $Y_B^-$ where the SFs start and end. Note that since $\Theta_{MP}$ does not depend on $\theta_F$, $Y_B^+$ and $Y_B^-$ also do not depend on $\theta_F$ as shown in Fig. \ref{fig:difftheta} for the SF states. Comparison of $Y_B^+$ and $Y_B^-$ obtained from (\ref{eq:Y_B}) with the numerical evaluation of $\Theta_{MP}$ and with the approximate expression (\ref{eq:TMP3order}) are shown in Fig. \ref{fig:tails} in the interval of interest for $P_B$ between 2 and 10.

\begin{figure}[]
    \centering
    \includegraphics[width=0.5\textwidth]{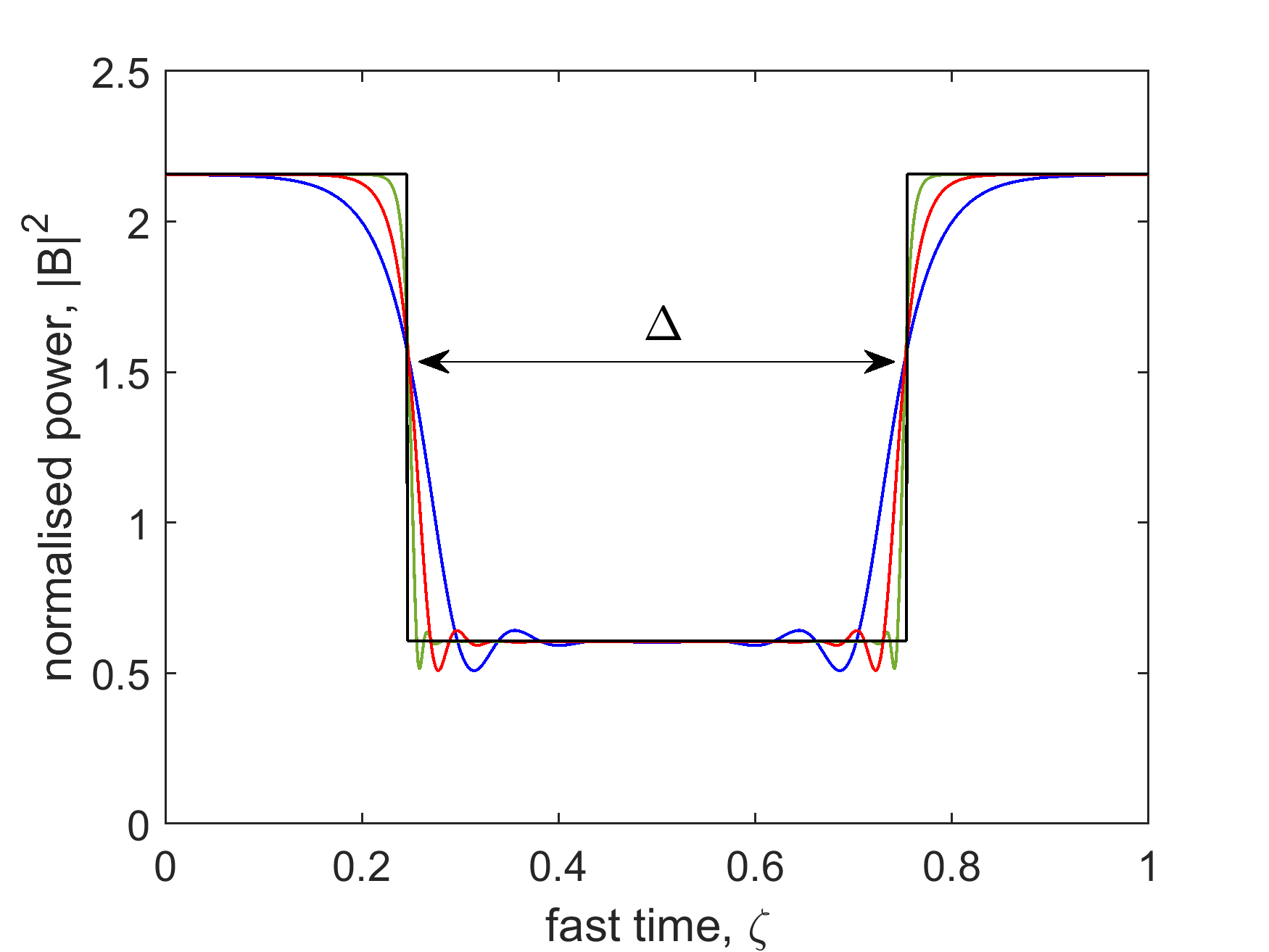}
    \caption{Power distribution of an inhomogeneous $B$ field exhibiting two non-interacting SFs with separation $\Delta$ for parameter values $P_F=P_B=2.1609$, $\theta_F = 1.4$, $\theta_B = 3.2$ and dispersion coefficient $\beta = 5$ (blue line), $\beta = 1$ (red line), $\beta = 0.1$ (green line), and $\beta = 0$ (black line). Here the fast time (x axis) is normalised to the round trip time.}
    \label{fig:fronts}
\end{figure}

It is now possible to obtain an expression for the stationary distance $\Delta$ of the two SFs. In the zero dispersion case $\beta=0$, the SFs are vertical lines between $Y_B^+$ and $Y_B^-$ so that 
\begin{eqnarray}
\langle |B|^2 \rangle &=& \Delta Y_B^- + (1-\Delta) Y_B^+ \nonumber \\
\Delta &=& \frac{Y_B^+ - \langle |B|^2 \rangle}{Y_B^+- Y_B^-} \label{eq:Delta}
\end{eqnarray}
However from (\ref{eq:SSforwards}) one obtains:
\begin{equation}
\langle |B|^2 \rangle = \frac{1}{\nu} \left[ \theta_F - |F|^2 \pm \sqrt{\frac{P_F}{|F|^2} -1} ~\right]
\label{eq:avB2}
\end{equation}
where $P_F$ is the forward input power $E_F^2$ and $|F|^2$ is given by Eq. (\ref{eq:Fsquare}). Hence the combinations of Eq. (\ref{eq:Y_B}) and Eq. (\ref{eq:avB2}) provide an expression of the distance $\Delta$ between the two SFs at zero dispersion via Eq. (\ref{eq:Delta}) in terms of parameters $\theta_F$, $\theta_B$, $P_F$, $P_B$ (see the black line in Fig. \ref{fig:fronts} for $P_F=P_B=2.1609$, $\theta_F = 1.4$, $\theta_B = 3.2$). For dispersion different from zero, the distance $\Delta$ remains unchanged as shown in Fig. \ref{fig:fronts} for $\beta = 5$ (blue line), $\beta = 1$ (red line), $\beta = 0.1$ (green line).

\begin{figure}[]
    \centering
    \includegraphics[width=0.5\textwidth]{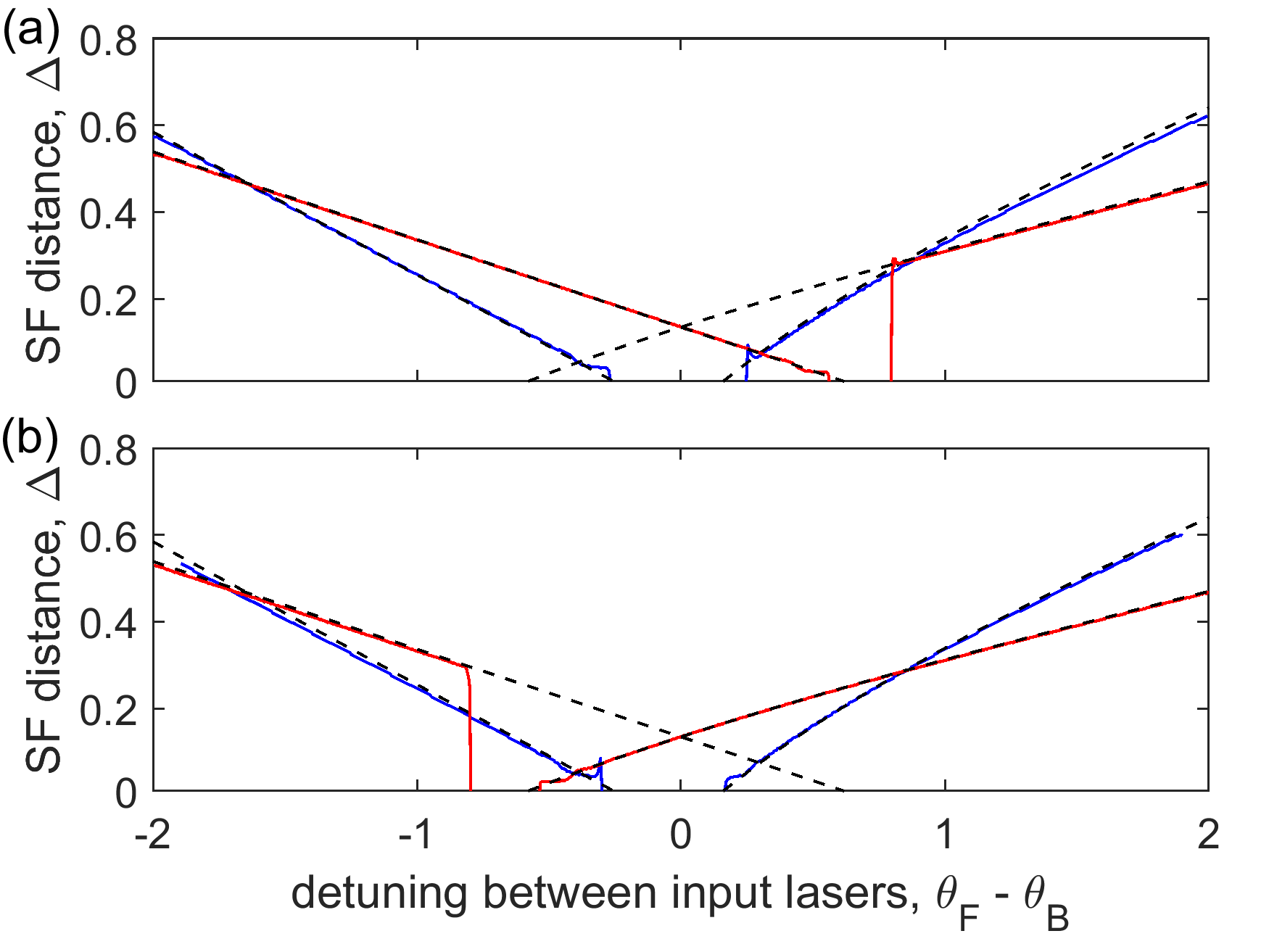}
    
    \includegraphics[width=0.25\textwidth]{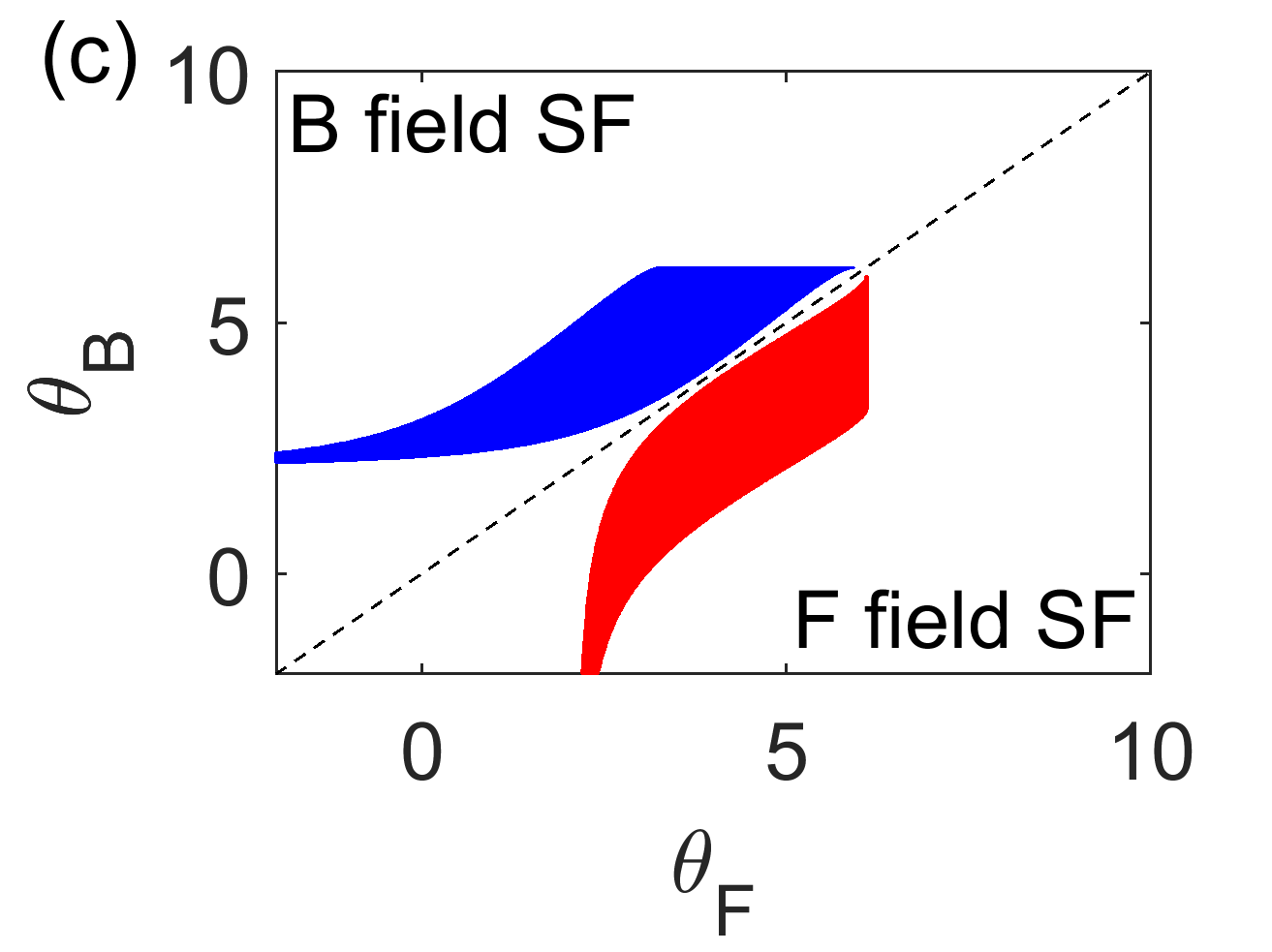}\includegraphics[width=0.25\textwidth]{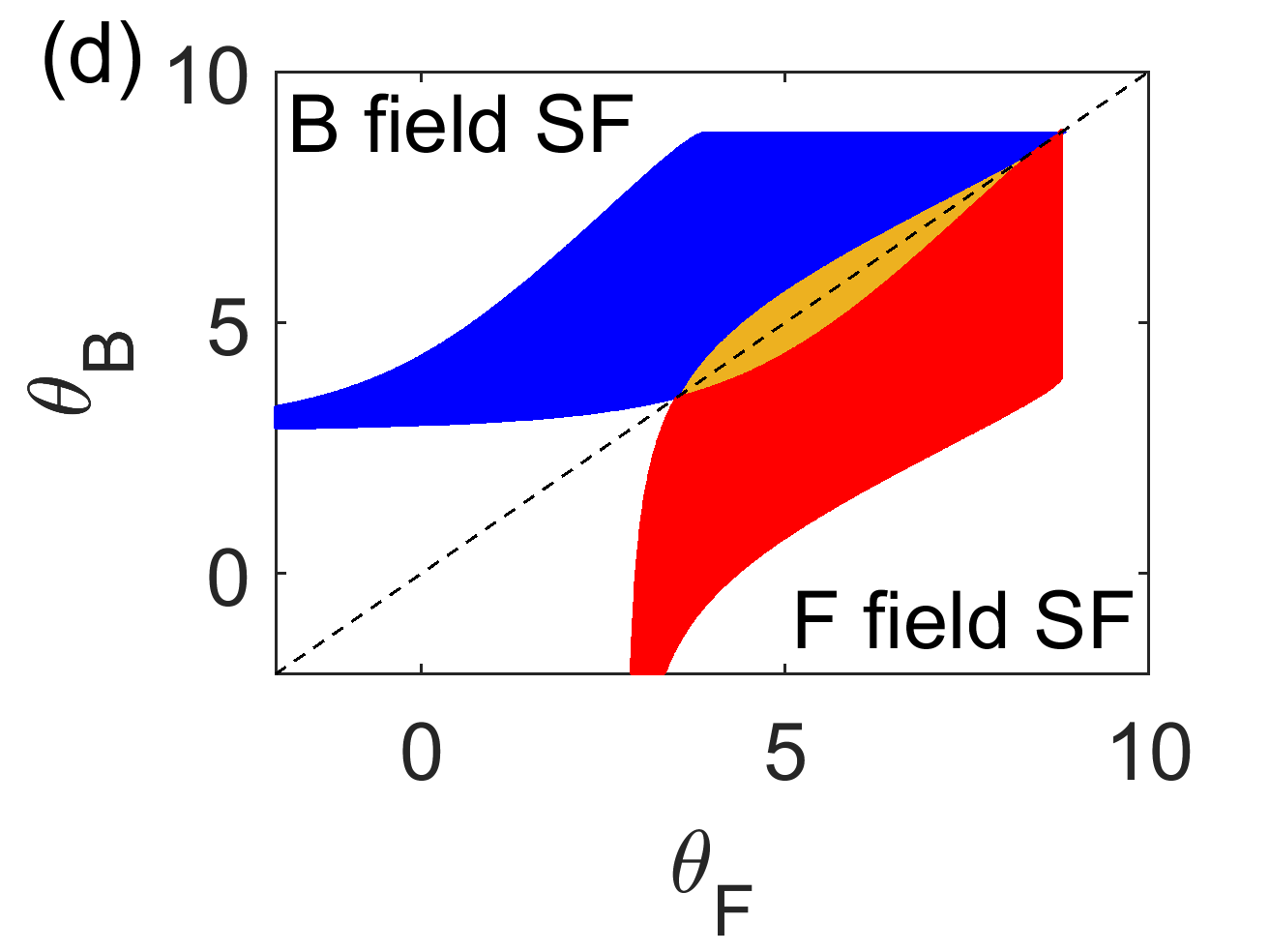}
    \caption{(a)-(b) SF separation $\Delta$ when changing the detuning $\theta_F$ for fixed pump powers $P_F=P_B$ and detuning $\theta_B$. Solid blue (solid red) lines correspond to simulation results from (\ref{eq:forwards})-(\ref{eq:backwards}) with $\beta=1$ and for $P_F=P_B=2.1609$, $\theta_B=3.2$ ($P_F=P_B=3,\theta_B=5$). (a) is a forward scan and (b) is a backwards scan. The black dashed lines are the analytical results of Eqs. (\ref{eq:Delta}) and (\ref{eq:Delta2}). (c)-(d) Range of detuning values where SF solutions exist and are stable for the $B$ field (blue region) and for the $F$ field (red region) or both fields (orange region), (c) $P_F=P_B=2$ and (d) $P_F=P_B=3$.}
    \label{fig:Delta}
\end{figure}

When using $\theta_F$ as a control parameter, expression (\ref{eq:Delta}) works very well when compared with the distance of two stationary SFs obtained from the simulations of (\ref{eq:forwards})-(\ref{eq:backwards}) done with $\beta=1$, see left hand side of Figs. \ref{fig:Delta}(a)-(b). In particular we note that $\Delta$ is a function of $\theta_F$ only through $\langle |B|^2 \rangle$ as expressed in Eq. (\ref{eq:avB2}). This means that the distance $\Delta$ decreases linearly with $\theta_F$ with a slope given by $[\nu (Y_B^+- Y_B^-)]^2$. Once the detuning $\theta_F<\theta_B$ and the input powers $P_B$ and $P_F$ are chosen, it is possible to obtain accurately the distance of the two SFs from Eq. (\ref{eq:Delta}) even in the regime of small distances and locked SFs (dark solitons) as shown in Fig. \ref{fig:Delta}.

The conditions of validity of Eqs. (\ref{eq:Fsquare}) and (\ref{eq:avB2}) predict that two stable SFs can be found in the interval $\eta (1+P_B)<\theta_B<\nu P_F + \eta (1+P_B)$, given that $0<\Delta<1$. This allows us to determine regions in parameter space where vertical SF form as shown in \ref{fig:Delta}(c)-(d). It is interesting to see that for values of $P_B>2.145$ where stable SFs in the backward field are observed even for $\theta_F>\theta_B$, the predictions of Eq. (\ref{eq:Delta}) remain in good agreement with the numerical results (see red lines on the left of Fig. \ref{fig:Delta} (a)-(b) for $P_B=P_F=3$).  

Eqs. (\ref{eq:Fsquare}) and (\ref{eq:Delta}) suggest that precise control over the pulse duration (SF distance) of the output field is possible by simply changing the laser detuning. This allows for control over the frequency comb generation efficiency by laser parameters in contrast with conventional micro-resonator dark solitons, where the pulse duration is determined by the dispersion.

We now move to the case $\theta_F>\theta_B$. In this case it is the backward field $B$ that is homogeneous and the two stable SFs are found in the forward field $F$. In this case the role of Eqs. (\ref{eq:SSforwards})-(\ref{eq:SSbackwards}) is exchanged:
\begin{eqnarray}
    E_F &=& (1 + i\Tilde{\theta}_F) F - i|F|^2 F + i \beta \partial_\zeta^2 F \label{eq:SSforwards2}\\
    E_B &=& (1 + i\Tilde{\theta}_B) B - i|B|^2 B \label{eq:SSbackwards2}
\end{eqnarray}
and one obtains $\Theta'_{MP} \approx \eta (1+P_F)$ as well as:
\begin{align}
    \langle |B|^2\rangle = |B|^2 
    = \frac{1}{\nu}[\theta_F -  \Theta'_{MP}] \approx \frac{1}{\nu}[\theta_F - \eta (1+P_F)]
    \label{eq:Bsquare}
\end{align}
In the case of $\theta_F>\theta_B$, the homogeneous power of the backward field grows linearly with $\theta_F$, which agrees with the simulation in Fig. \ref{fig:difftheta}b. The form of the equation for the power $Y_F=|F|^2$, however, remains basically unchanged from Eq. (\ref{eq:Y_B}),
\begin{align}
Y_F^3-2 \Theta'_{MP} Y_F^2 + \left( 1+(\Theta'_{MP})^2 \right) Y_F - P_F = 0
\label{eq:Y_F}
\end{align} 
so that the homogeneous powers $Y_F^+$ and $Y_F^-$ before and after the SFs are still independent from $\theta_F$ and, in the case of $P_F=P_B$, they have the same values of $Y_B^+$ and $Y_B^-$ found for $\theta_F<\theta_B$ since $\Theta'_{MP}=\Theta_{MP}$. Finally,
\begin{eqnarray}
\langle |F|^2 \rangle &=& \Delta Y_F^- + (1-\Delta) Y_F^+ \nonumber \\
\Delta &=& \frac{Y_F^+ - \langle |F|^2 \rangle}{Y_F^+- Y_F^-} \label{eq:Delta2}
\end{eqnarray}
and
\begin{equation}
\langle |F|^2 \rangle = \frac{1}{\nu} \left[ \theta_B - |B|^2 \pm \sqrt{\frac{P_B}{|B|^2} -1} ~\right]
\label{eq:avF2}
\end{equation}
The distance $\Delta$ depends on $\theta_F$ through $\langle |F|^2 \rangle$ and then through $|B|^2$ given in Eq. (\ref{eq:Bsquare}) and (\ref{eq:avF2}). At difference from the case $\theta_F<\theta_B$ this dependence is nonlinear, the slope of the curve is reversed and the distance $\Delta$ now grows with the detuning $\theta_F$. The agreement of Eq. (\ref{eq:Delta2}) with the numerical simulations as shown in the right hand part of Fig. \ref{fig:Delta} is again excellent. Similar to the $B$ field case, the conditions of existence of vertical SFs for the $F$ field is $\eta (1+P_F)<\theta_F<\nu P_F + \eta (1+P_F)$ given that $0<\Delta<1$ (see Fig. \ref{fig:Delta}c-d).

The linear stability of SF solutions can be determined at zero dispersion using the expressions for the average field powers derived earlier in this section. Considering a SF solution in the backward field with a homogeneous forward field, their average powers are given by Eqs. (\ref{eq:avB2}) and (\ref{eq:Y_F}), respectively. As calculated in Appendix \ref{app:linear_stability_inhomo} the stability of the homogeneous states before and after the SFs to spatial (fast time) perturbation are given by the eigenvalues 
\begin{eqnarray}
    \lambda_B^+ &=& -1 \pm \sqrt{(\Theta_{MP}-Y_B^+)(3Y_B^+-\Theta_{MP})}\\
    \lambda_B^- &=& -1 \pm \sqrt{(\Theta_{MP}-Y_B^-)(3Y_B^--\Theta_{MP})} \, .
\end{eqnarray}
These eigenvalues depend on the pump power only. When changing the detuning $\theta_F$, the corresponding SF solution maps into one of the multi-stable two SF solutions of an LLE at Maxwell point. The homogeneous forward field eigenvalues are
\begin{eqnarray}
    \lambda_F &=& -1 \pm \sqrt{(\Tilde{\theta}_F-Y_F)(3Y_F-\Tilde{\theta}_F)}
\end{eqnarray}
where $\Tilde{\theta}_F = \theta_F - \nu\langle|B_s|^2\rangle$ is the effective detuning, and depend on $\theta_B$ implicitly through the integrated power $\langle |B|^2\rangle$. 

By using the stability eigenvalues $\lambda_B^\pm$ and $\lambda_F$ it is possible to determine instabilities of the SF solutions when the real part of one of these eigenvalues goes from negative to positive. For example plateau solutions separated by SF are susceptible to Hopf bifurcations and oscillations of the homogeneous states that are connected to the SFs. This instability is introduced by perturbations to the SF states that change the average power of the field as seen in Appendix \ref{app:linear_stability_inhomo}. For the parameter values used in this work $P_F=P_B=2.1609,\theta_B=3.2$, these oscillations grow in the region $5.35<\theta_F<6.25$ resulting in the collapse of local structures to the HSS. Numerical simulations of Eqs. (\ref{eq:forwards})-(\ref{eq:backwards}) confirming this instability are presented in Section VI. 


\section{Evolution towards the two switching-front solutions}\label{sec:soldy}
Despite the one to one correspondence of the SF solutions of the counterpropagating system and those of the LLE at Maxwell point, the dynamics of front solutions in the counterpropagating system are different form those seen in the LLE. Here we describe first the transient evolutions of a two SF solution in the counterpropagating system as the SFs move towards the unique stationary separation of the fronts.
\begin{figure*}[]
    \centering
    {\includegraphics[width=0.32\textwidth]{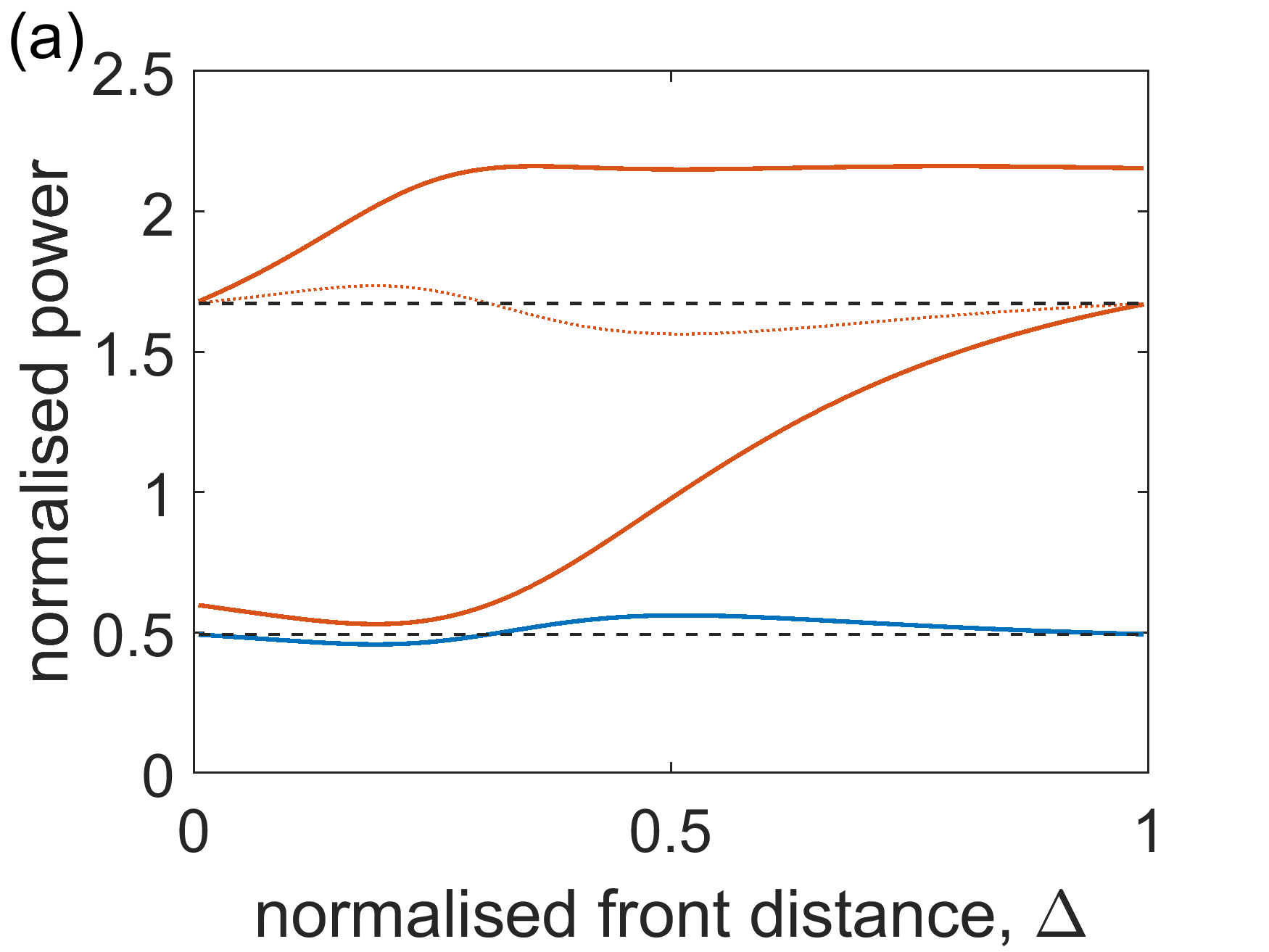}}
    {\includegraphics[width=0.32\textwidth]{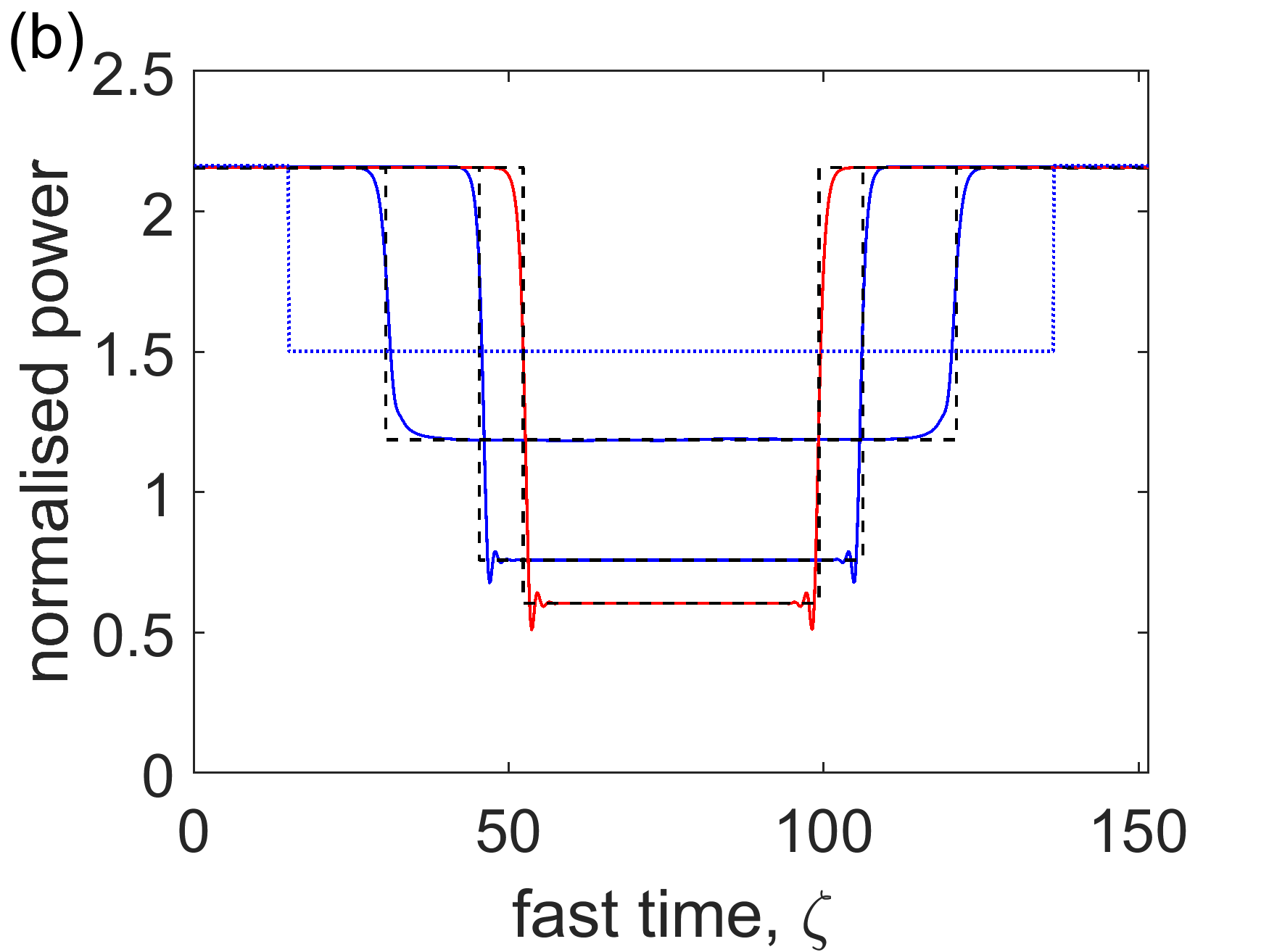}}
    {\includegraphics[width=0.32\textwidth]{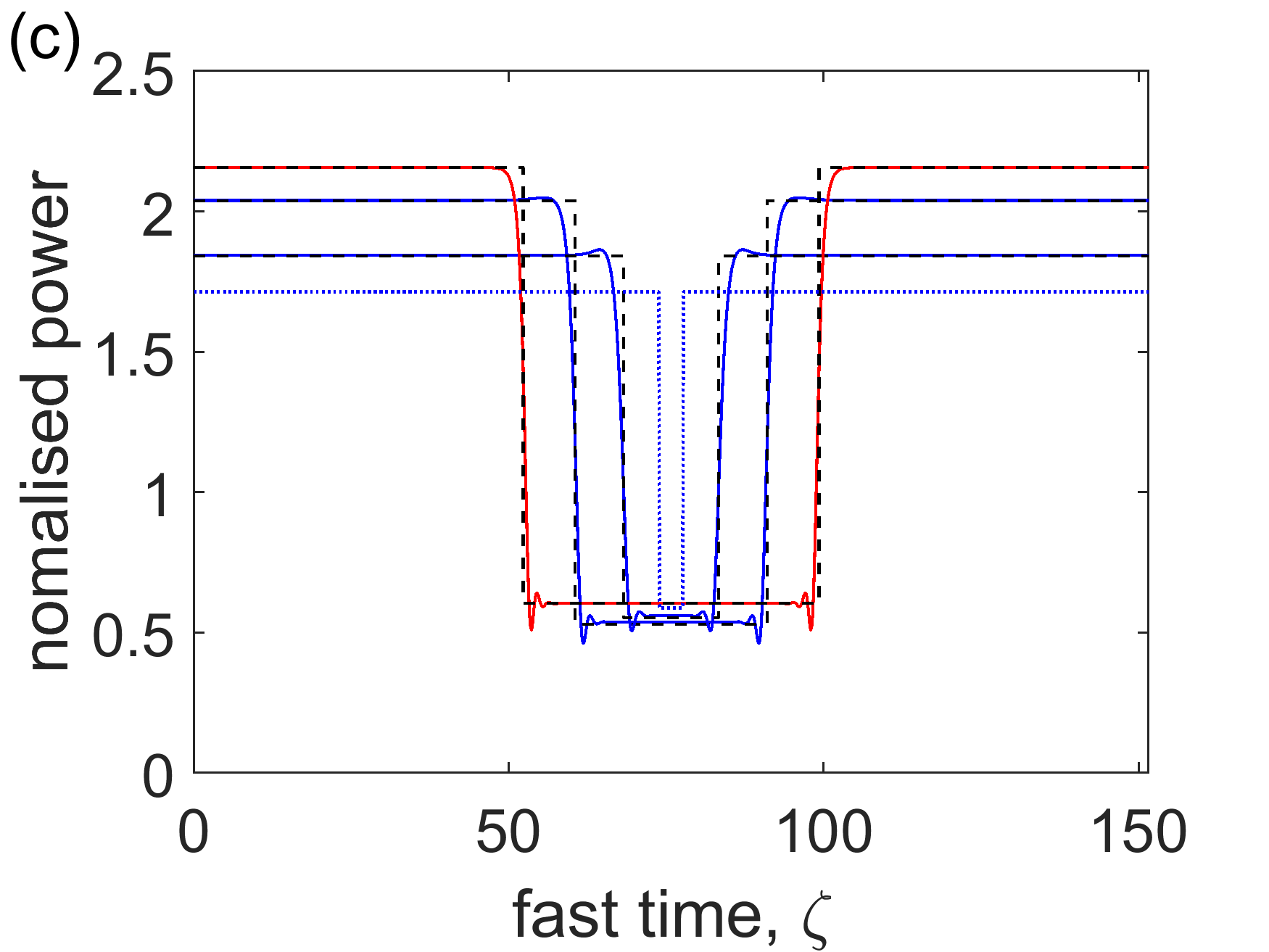}}
    \caption{(a) Power of the homogeneous states $Y_B^\pm$ connected by the SFs (solid red line), homogeneous field power $H_F$ (solid blue line), and average power of the field displaying SFs (dotted red line) versus the front separation is changed. The HSS in the absence of SFs is given by the dashed black line. (b)-(c) Comparison between the zero dispersion two front solutions using Eqs. (\ref{eq:shiftedHSS})-(\ref{eq:uppersplitHSS}) (black dashed lines) and evolving two front solutions from the numerical integration of Eqs. (\ref{eq:SSforwards})-(\ref{eq:SSbackwards}) with $\beta=0.1$ (solid lines) for shrinking front distance (b) and expanding front distance (c). The dotted lines are the initial conditions. Parameter values are $P_F=P_B=2.1609$, $\theta_F=2.0$, and $\theta_B=3.2$. The final and stationary front separation (red line) is $\Delta=0.31$ in both (b)-(c).}
    \label{fig:solitonevolutionICsol}
\end{figure*}

In Fig. \ref{fig:movingfronts} in Section III, we have seen that when the HSS of the counterpropagating system are unstable to inhomogeneous perturbations, the system relaxes to a SF solution. We consider here initial conditions made of two SFs between two homogeneous states in one field (the backward one for $\theta_F<\theta_B$) while the other field is homogeneous across the resonator. When the front separation is not at the stationary value, the values of the homogeneous states at the beginning and at the end of each front in the counterpropagating system depend on the average power of the fields. This means that these values are different from those at the final front separation at the stationary value. The values of the homogeneous power before and after a front for arbitrary separations can be calculated by considering states of the zero dispersion case of Eqs. (\ref{eq:forwards})-(\ref{eq:backwards}), where the second order derivative with respect to the fast time and the first derivative with respect to the slow time are neglected. For a two front solution in the $B$ field, the upper and lower homogeneous solutions separating the SFs can be determined by solving the coupled equations
\begin{eqnarray}
    P_B &=& Y_B^3 - (\theta_B - \nu Y_F)Y_B^2 +[(\theta_F - \nu Y_F)^2+1]Y_B\label{eq:shiftedHSS} \\
    P_F &=& Y_F^3 - (\theta_F - \nu [\Delta Y_B^- + (1-\Delta)Y_B^+])Y_F^2 \nonumber \\
        &\quad&\quad+[(\theta_F - \nu [\Delta Y_B^- + (1-\Delta)Y_B^+])^2+1]Y_F 
\label{eq:uppersplitHSS} 
\end{eqnarray}
where $Y_B^+,Y_B^-$ are the upper and lower homogeneous solutions of the zero dispersion SF solution present in the $B$ field (solutions of Eq. (\ref{eq:uppersplitHSS}) in a bistable state) with average power $\langle|B|^2\rangle = \Delta Y_B^-(\Delta)+(1-\Delta)Y_B^+(\Delta)$ and $\Delta$ is the front separation. Note that the expressions for the average powers of front solutions are independent of dispersion. These solutions are plotted in Fig. \ref{fig:solitonevolutionICsol}a. 
%
\begin{figure}[]
    \centering
    {\includegraphics[width=0.23\textwidth]{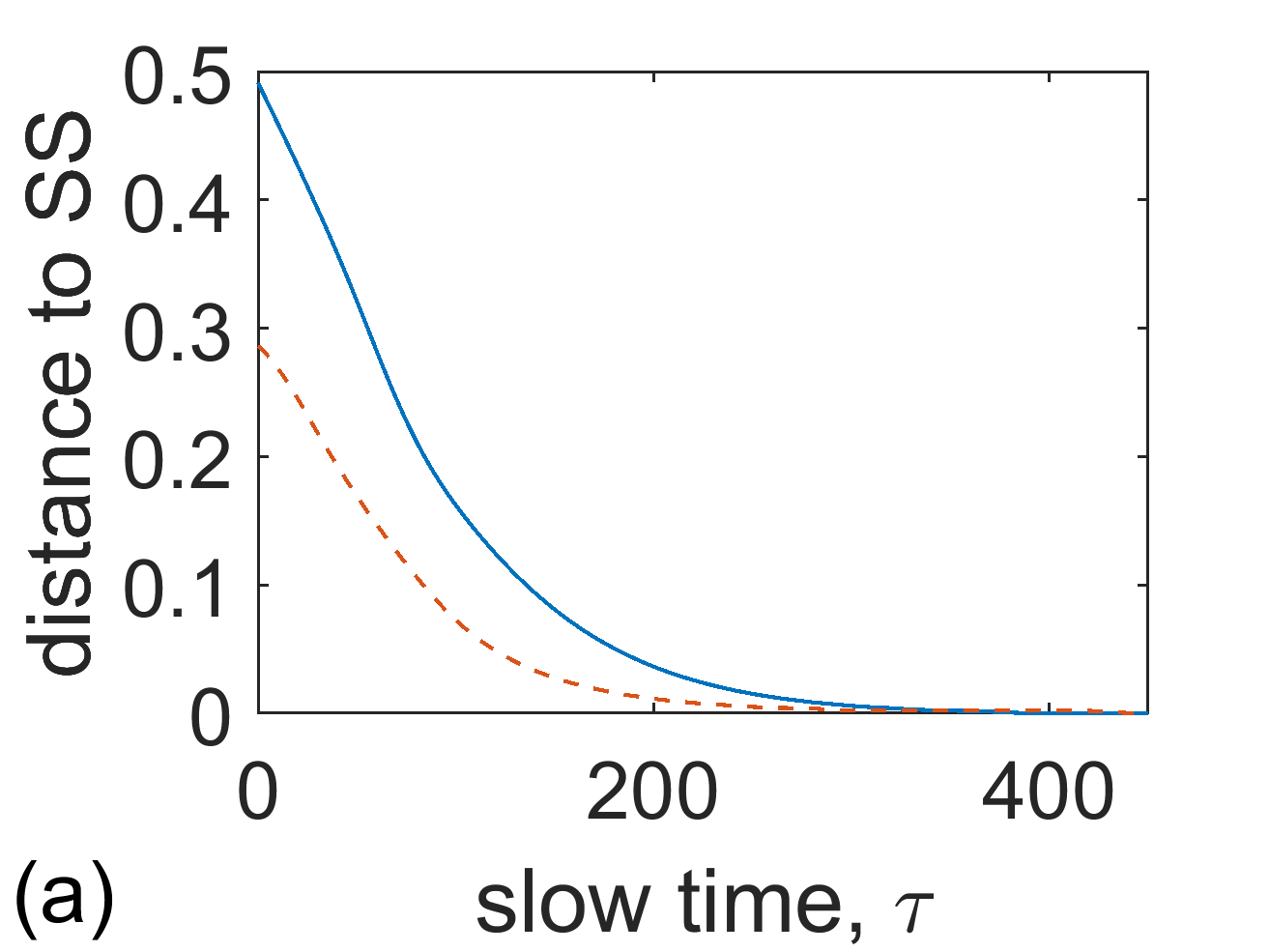}}
    {\includegraphics[width=0.23\textwidth]{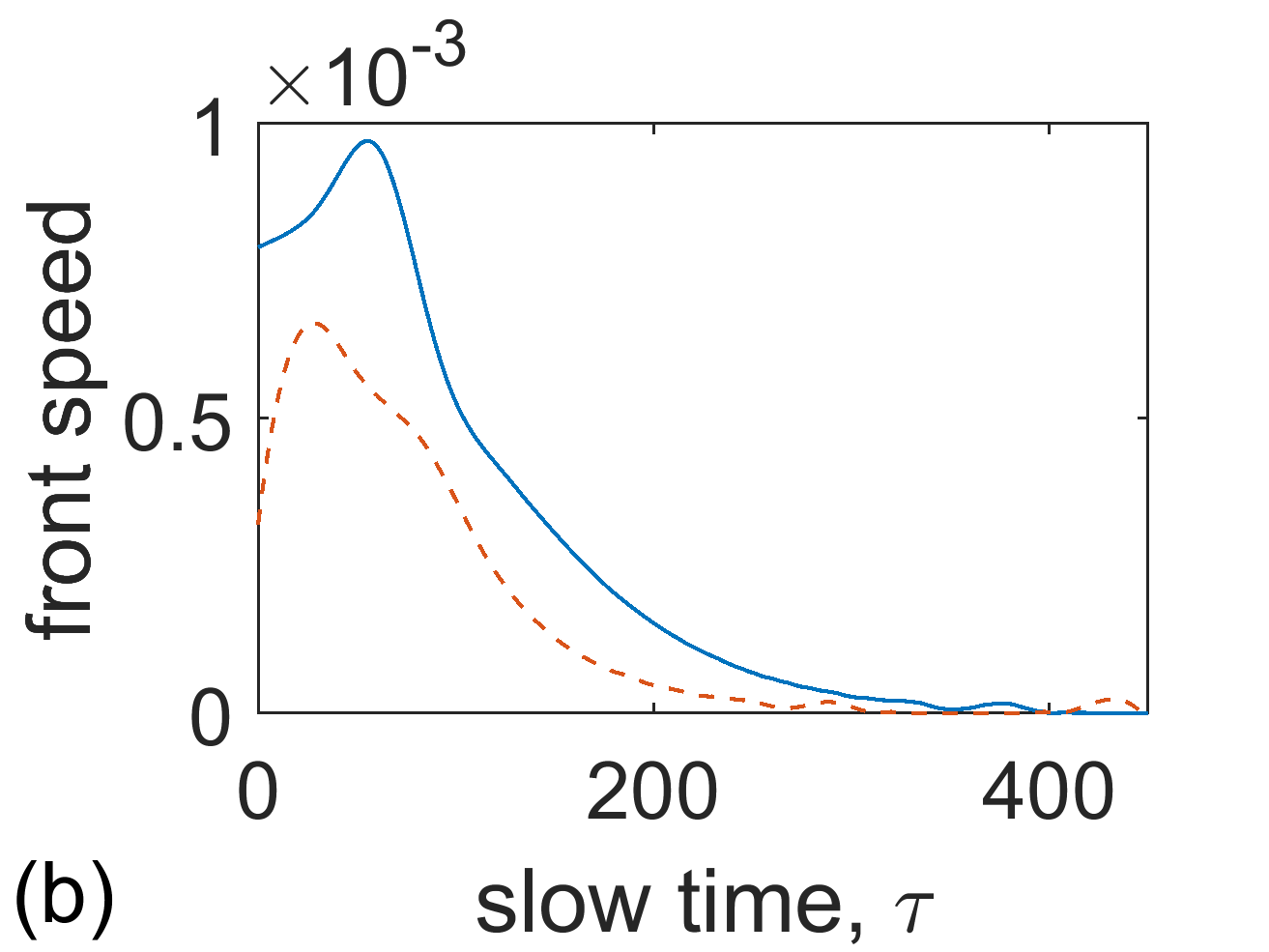}}
    \caption{Front separation (a) and front velocity (b) vs slow time while approaching a SF solution. From the data from Fig. \ref{fig:movingfronts}, we track the front separation relative to the separation of the final SF solution in (a), use the dimensionless slope of (a) to determine the front speed in (b). Solid blue line represents the wide initial condition, red dashed line the narrow initial condition.}
    \label{fig:sepvelvstime}
\end{figure}

Fig. \ref{fig:solitonevolutionICsol}b and c show that two-front profiles that use the solutions of Eqs. (\ref{eq:shiftedHSS})-(\ref{eq:uppersplitHSS}) with a given separation $\Delta$ provide excellent approximations to the numerical solutions of Eqs. (\ref{eq:SSforwards}) and (\ref{eq:SSbackwards}) with $\beta=1$ during the transients to the the final SF solution for both cases of shrinking and expanding front separation. The SFs are moving with opposite velocities and with a well defined distance $\Delta(t)$. For each value of the slow time $t$ and distance $\Delta(t)$, the dynamical solution is well approximated by two SFs between homogeneous states provided by Eqs. (\ref{eq:uppersplitHSS}) given a separation distance $\Delta$. Since for each value of $\theta_F$ there is only one stationary value of $\Delta$, generic separations of the two SFs separated by homogeneous power from Eqs. (\ref{eq:uppersplitHSS}) evolve in time but maintain their shape with a changing separation leading to different homogeneous powers. As such the front separation determines the power of homogeneous solutions, which in turn determines the velocity of the SFs, which in turns changes the front separation. This leads to a front velocity that depends on the front separation.

Although the shape of the transient solutions are well approximated by two vertical SFs at every moment in time, the front separation and the front velocity are non-trivial functions of time as shown in Fig. \ref{fig:sepvelvstime}.

\section{oscillatory dynamics and bistability with front stationary states}\label{sec:Oscill}
\begin{figure}[]
    \centering
    {\includegraphics[width=0.25\textwidth]{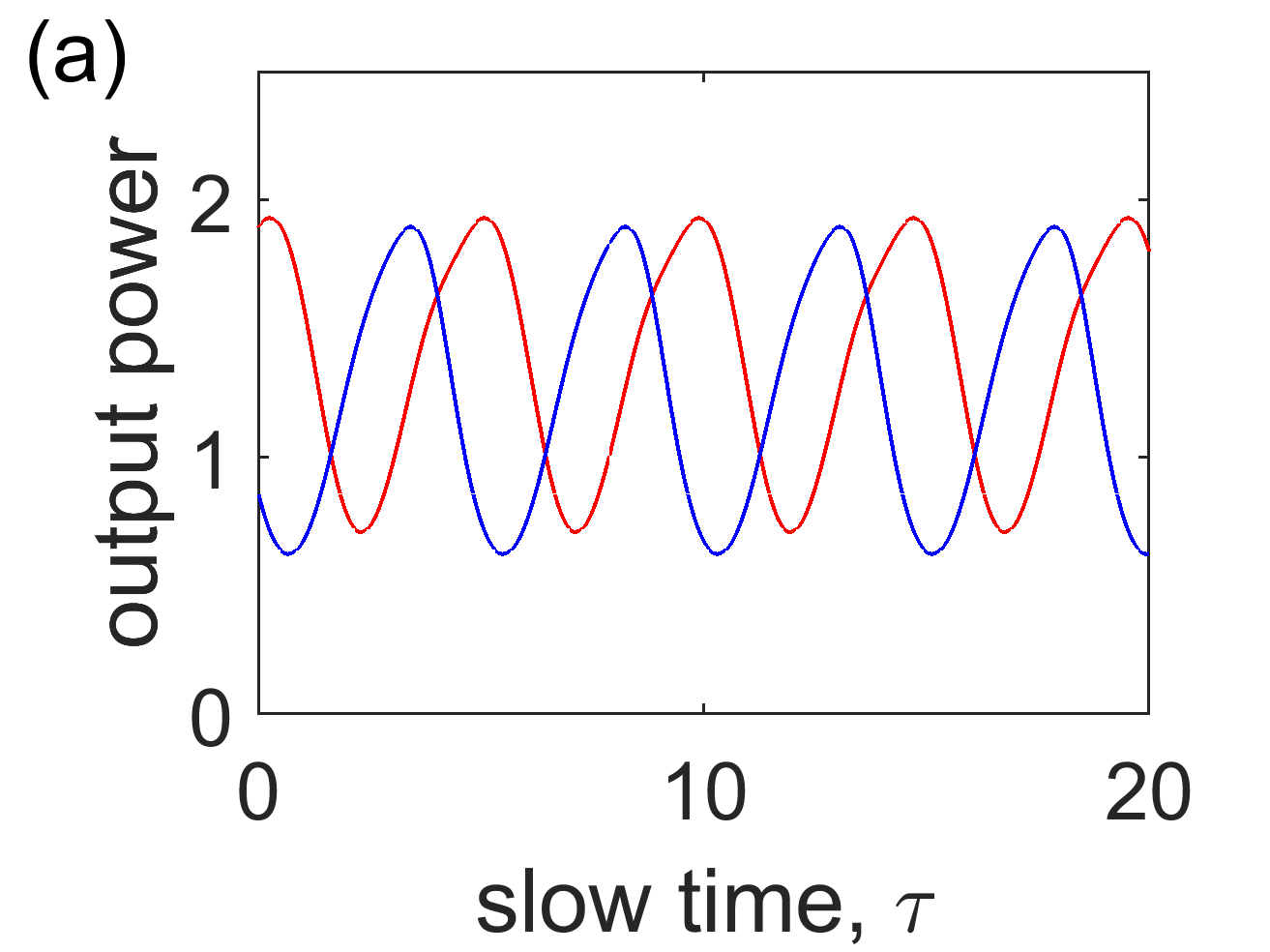}}{\includegraphics[width=0.25\textwidth]{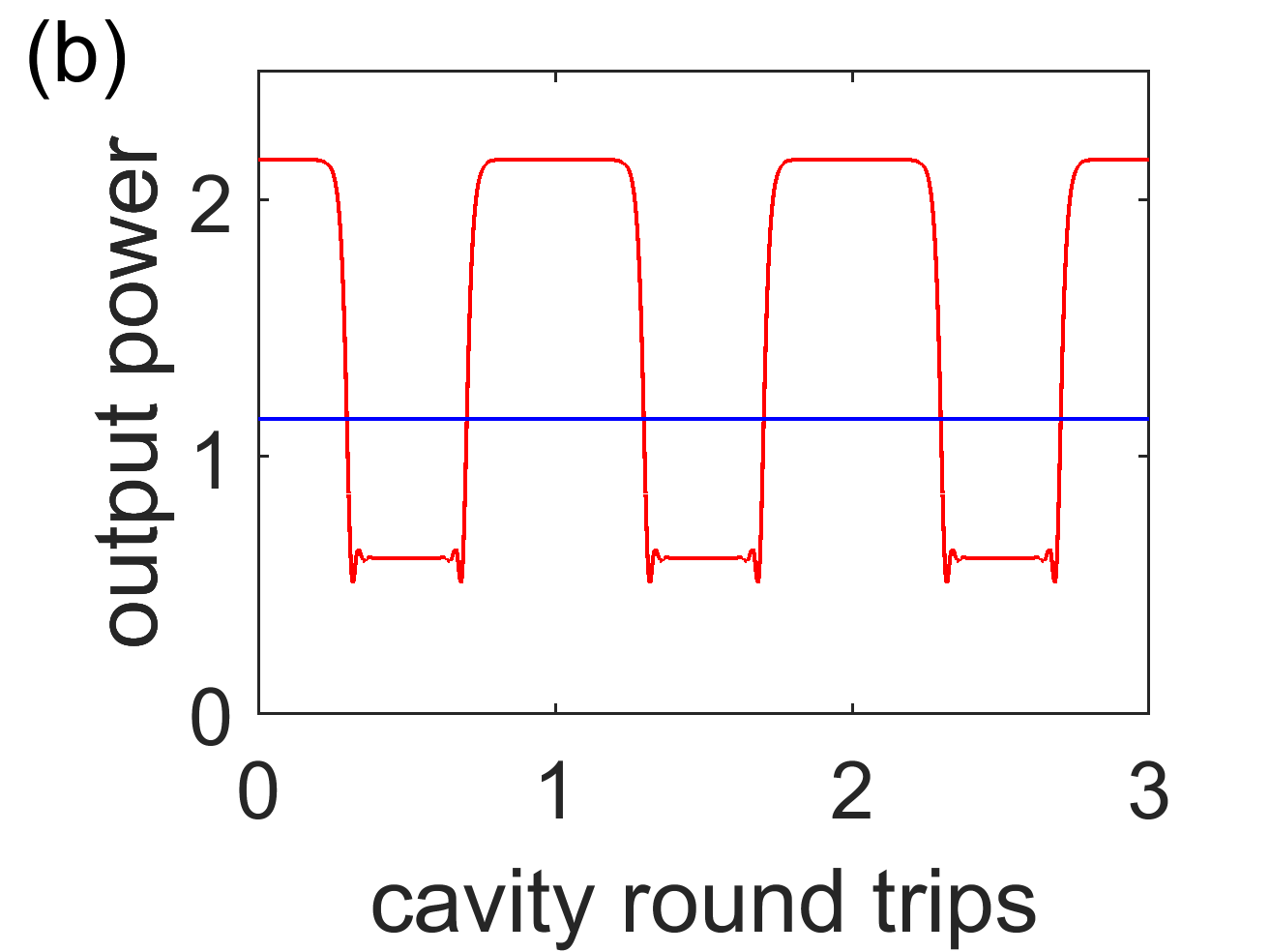}}
    \caption{Bistability of slow and fast oscillations for parameter values $\beta=1$, $P_F=P_B=2.1609$, $\theta_F=4.5$, and $\theta_B=3.2$. (a) Periodic oscillations of the homogeneous powers of both counterpropagating fields over the slow time. (b) Output power of a SF solution in the forward field and homogeneous steady state for the backward field over three cavity round trip times.}
    \label{fig:oscillations}
\end{figure}

Dynamical regimes in ring resonators have been previously studied for homogeneous counterpropagating fields with symmetrical input fields and detunings \cite{woodley2018universal,woodleyPRL21}. It was seen that under the correct conditions, a pair of oppositely directed Hopf bifurcations can occur when changing the detuning $\theta_F=\theta_B$, allowing for sustained homogeneous oscillations that could exhibit period doubling bifurcations, chaos and crisis events. In Section III we saw oppositely directed Hopf bifurcation for the HSS occurring when changing $\theta_F$ in an asymmetric regime of different detunings between the two counterpropagating waves since $\theta_B$ is kept fixed (see the vertical black dashed lines in Fig. \ref{fig:HSS} evidencing the interval $4.02<\theta_F<6.33$). These Hopf bifurcations affect the highest power HSS resulting in oscillations which are bistable with the lowest power HSS. An example of large homogeneous oscillations in the power of the two fields is displayed in Fig. \ref{fig:oscillations}a from simulations of Eqs. (\ref{eq:forwards})-(\ref{eq:backwards}). 

\begin{figure}[]
    \centering
    \includegraphics[width=0.5\textwidth]{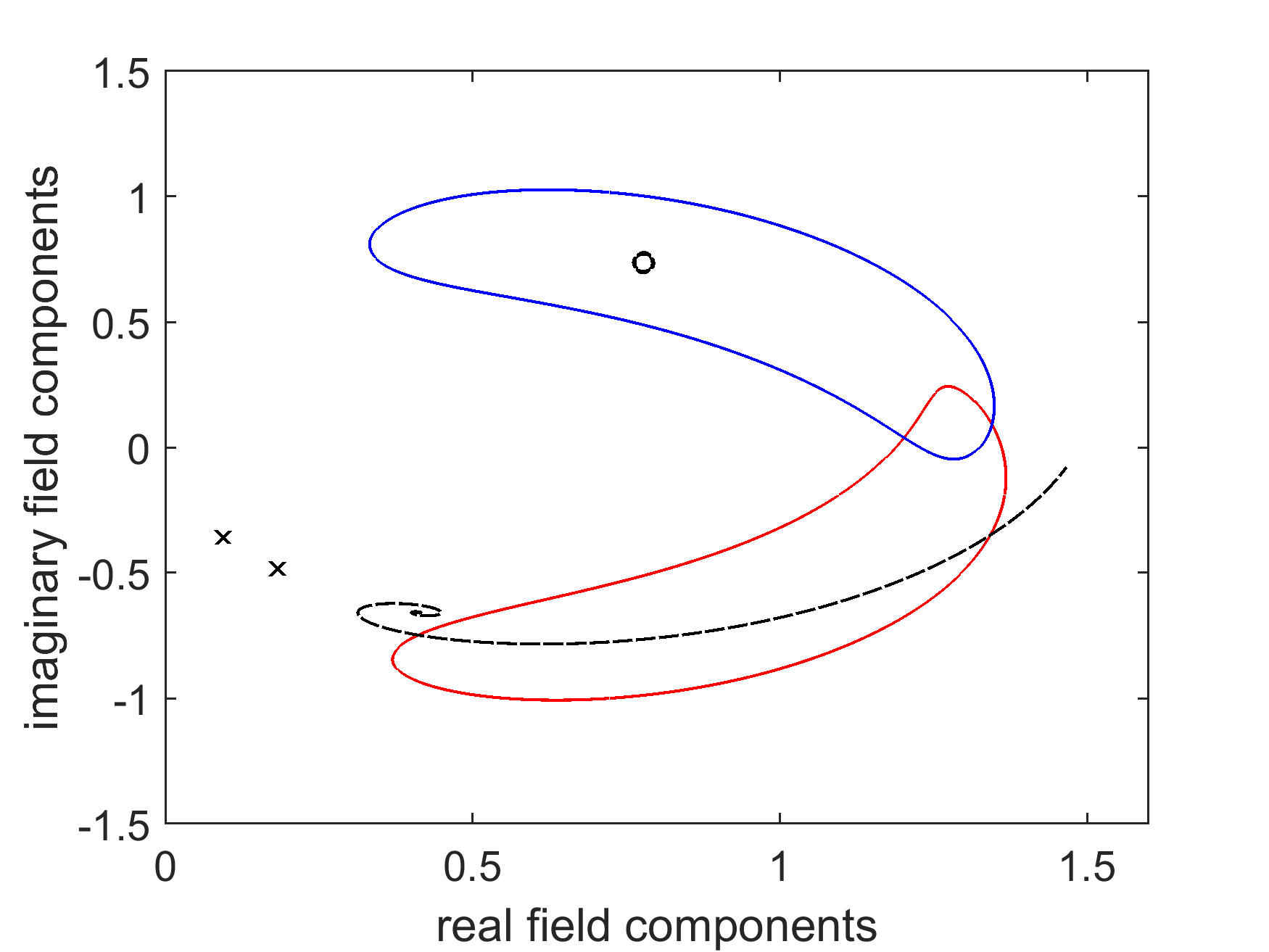}
    \caption{Possible asymptotic states for $\beta=1$, $P_F=P_B=2.1609$, $\theta_F=4.5$, and $\theta_B=3.2$ in the phase (Argand) plane. Stable limit cycle trajectories of the homogeneous forward (red solid line) and backward (blue solid line) fields; stable SF solution of the forward field (black dashed line) and its homogeneous backward field (black circle); stable HSS of low powers (black Xs for forward and backward fields).}
    \label{fig:bistability_TFSS_HSdynamics}
\end{figure}

In the parameter region of Fig. \ref{fig:oscillations}, the HSS of large powers are unstable not only to homogeneous oscillations but also to local perturbations on the fast time scale (see the line marked with X in the interval $3.35<\theta_F<6.47$ in Fig. \ref{fig:HSS}). We find that depending on the initial condition, the system can evolve to either the homogeneous oscillations of Fig. \ref{fig:oscillations}a or to a SF solution in the forward field with a homogeneous backward field (see Fig. \ref{fig:oscillations}b) or to a HSS corresponding to low powers. To display the richness of possible asymptotic states of Eqs. (\ref{eq:forwards})-(\ref{eq:backwards}), we show in Fig. \ref{fig:bistability_TFSS_HSdynamics} the asymptotic trajectories of oscillating homogeneous fields, the asymptotic trajectories of the SF state and the asymptotic points of the HSS of low powers in the phase (Argand) plane for the same parameters of Fig. \ref{fig:oscillations}.
Depending on the initial condition, the micro-ring device can evolve to any of these three final states generating either large amplitude slow oscillations in both fields, or large amplitude fast oscillations in just one field (the forward one) or no output oscillations at all. This provides the operator with a remarkable number of output waveforms with possible selection of each one by suitable perturbation of the input fields (in their amplitude or phase). 
\begin{figure}[]
    \centering
    \includegraphics[width=0.5\textwidth]{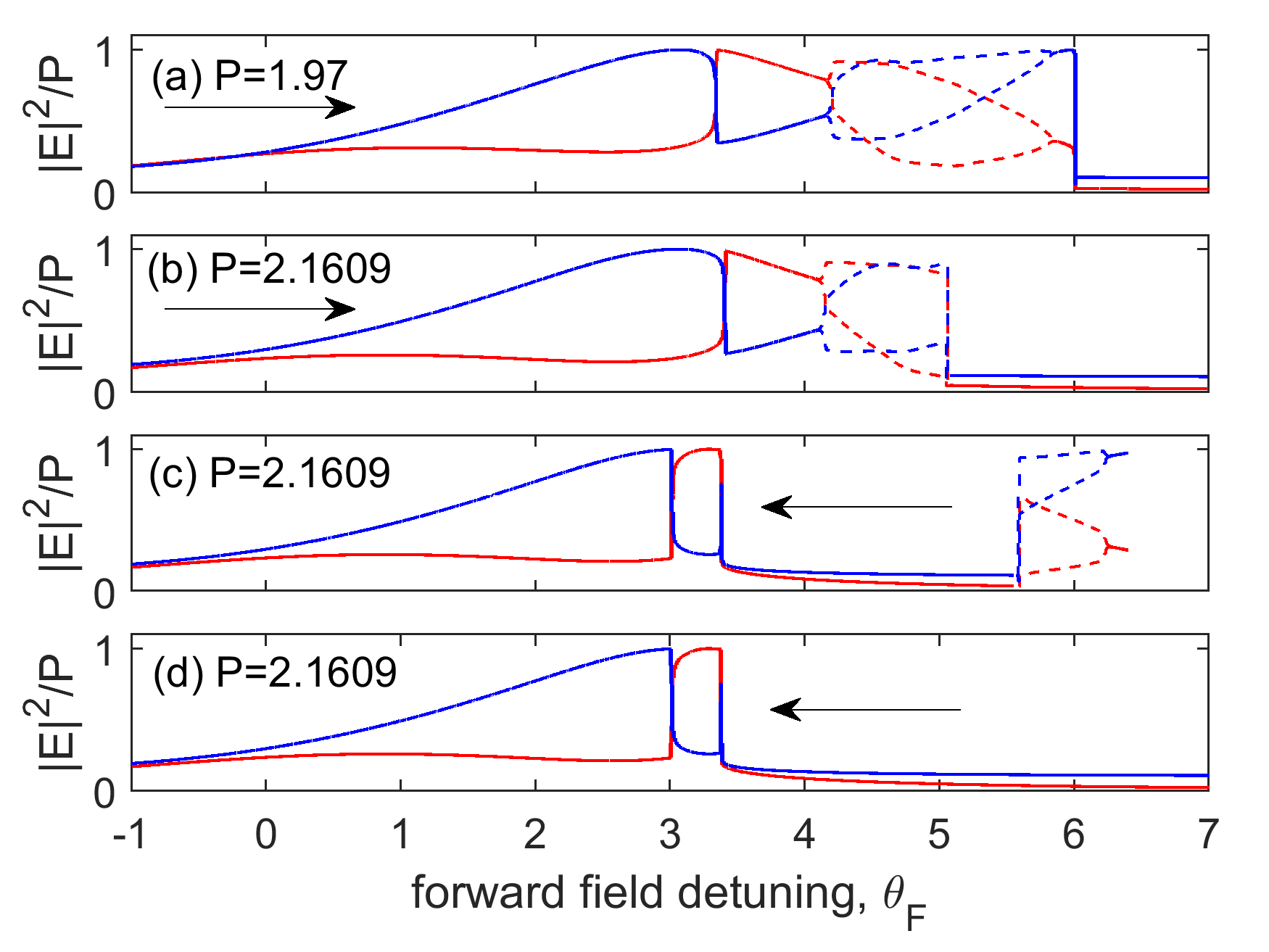}
    \caption{Homogeneous field powers when scanning the detuning $\theta_F$ for fixed detuning $\theta_B=3.2$ and fixed equal pump powers $P=P_F=P_B$. Dashed lines correspond to the power extrema during oscillation. 
    (a) Forward scan for $P=1.95$. Limit cycle oscillations are present in the detuning range $4.2<\theta_F<5.9$. (b) Forward scan for $P=2.1609$. Limit cycle oscillations are present in the detuning range $4.1<\theta_F<5.1$. (c) Backward scan for $P=2.1609$ starting at $\theta_F=6.4$. Limit cycle oscillations are present in the detuning range $5.5<\theta_F<6.2$. (d) Backward scan for $P=2.1609$ starting at $\theta_F=7.0$. No oscillations observed.  
    }
    \label{fig:dynamics_scan}
\end{figure}

When scanning the forward detuning for the parameter values studied here, we do not observe period doubling bifurcations or deterministic chaos at difference with typical simulations at parameter symmetry \cite{woodley2018universal,hill2020effects,woodleyPRL21}. 
We observe however sudden crises when the stable trajectory of the limit cycle can intersect the unstable HSS in the regions of multiple stationary states. This results in sudden instabilities of the oscillations, which collapse to the lower stable HSS. In Fig. \ref{fig:dynamics_scan} we show simulations of counterpropagating fields when scanning the detuning $\theta_F$ forwards and backwards. Forward and backward Hopf bifurcations can be clearly seen in the forward scan at $P_F=P_B=1.95$ in Fig. \ref{fig:dynamics_scan}a where the dotted lines represent the maxima and minima of the oscillating powers of the homogeneous fields over slow time variations. When increasing the input power, attractor crises are observed both in the forward (at $\theta_F\approx5.05$) and in the backward (at $\theta_F\approx5.62$) scans (see Fig. \ref{fig:dynamics_scan}b--c) leading to transfers to the low power HSS. Note however that depending on the initial condition of the backward scan, there is the possibility of observing no oscillations and no crises as displayed in Fig. \ref{fig:dynamics_scan}d.

\begin{figure}[]
    \centering
    \includegraphics[width=0.5\textwidth]{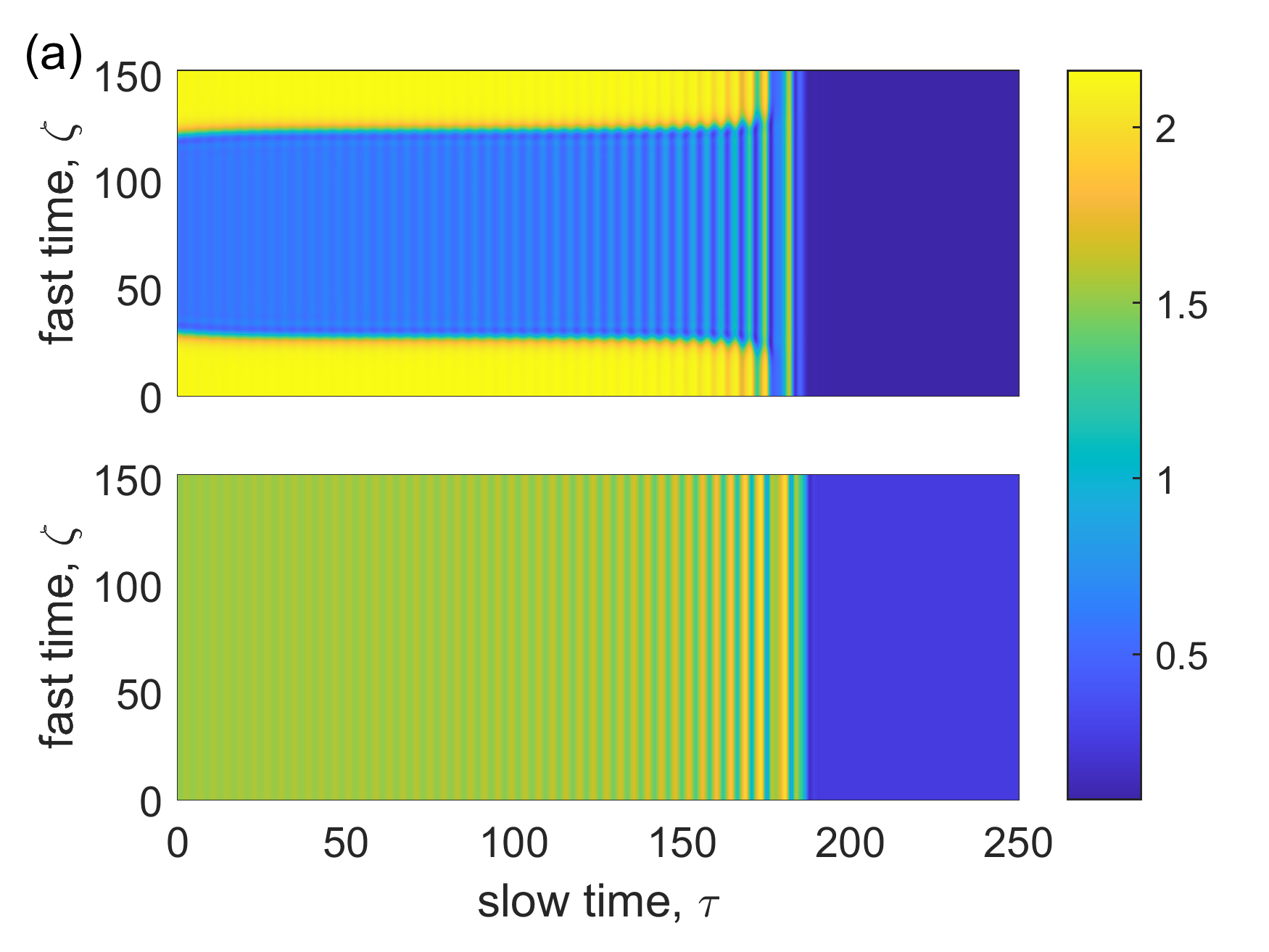}
    \includegraphics[width=0.5\textwidth]{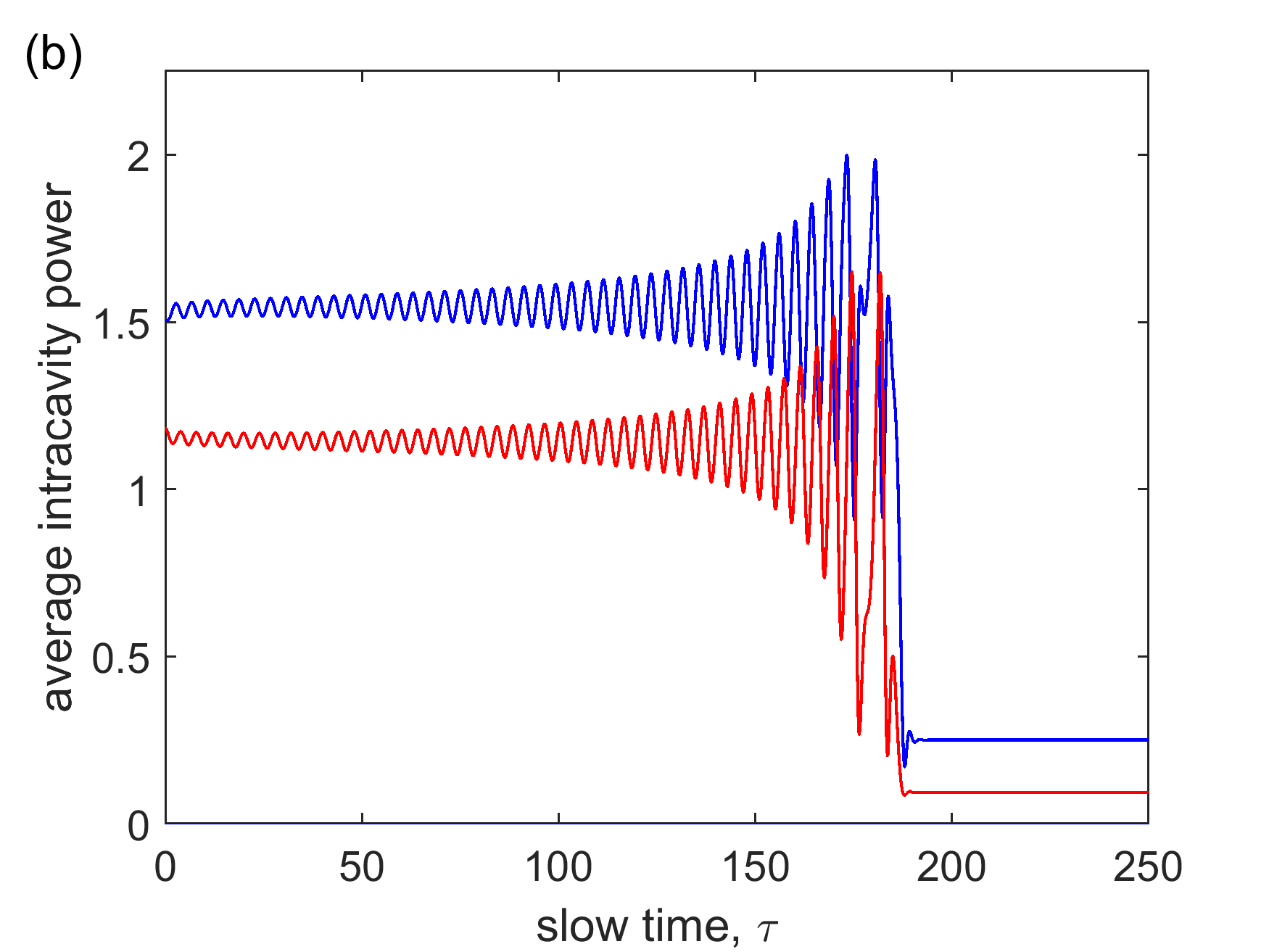}
    \caption{Dynamical evolution from an initial condition of a two SF solution in the forward field and a homogeneous solution in the backward field for $P_F=P_B=2.1609, \theta_F=5.3, \theta_B = 3.2$. Oscillations grow until both fields reach the stable HSS of low powers. (a) Intracavity power of the forward (upper) and backward (lower) fields over slow time. (b) Average interactivity power of the forward (red) and backward (blue) over slow time.}
    \label{fig:TFSS_instability}
\end{figure}

Fig. \ref{fig:dynamics_scan} focuses on homogeneous oscillations and HSS of low powers. The situation is further complicated by the presence of SF states in the forward field with a homogeneous backward field. When changing $\theta_F$ there is a further temporal instability of the SF solutions which causes the homogeneous states connecting the SFs to start to oscillate resulting in the entire inhomogeneous structure to oscillate, along with homogeneous oscillations of the backward field. For $\theta_F<5.35$ these oscillations are damped allowing for stable SF states, but for $5.35<\theta_F<6.25$ such oscillations grow, destroying fast time structures and the system moves to the HSS corresponding to low powers as shown in Fig. \ref{fig:TFSS_instability}.
\section{Conclusions}
We have investigated a nonlocally coupled model describing the interaction of two input beams counterpropagating in a ring resonator with normal dispersion. In particular, we have derived a semi-analytical description of plateau solutions separated by two switching fronts via a one to one correspondence with multi-stable SF at arbitrary separations exhibited by the Lugiato-Lefever equation at the Maxwell point. At difference from locally coupled LLEs, the distance of two stationary SFs in the counterpropagating case is controllable by the detunings which in turn can be tuned by changing the frequency of the input fields. Robust SF solutions are present for large ranges of detuning allowing great control over the distance of the two SFs through the laser parameters. This allows us to precisely control the pulse duration of the output field, and hence the frequency comb generation efficiency by changing the input laser detunings. This is different from conventional micro-resonator dark solitons, whose width is determined by the dispersion. In addition, our model predicts for low input field powers that changing the laser detuning across the symmetric state will result in the SF solutions to disappear from one field and then to reappear in the other field. This results in the SFs switching direction in the micro-resonator while scanning a single detuning parameter. The analytic description of SF solutions extends to the transient states too, allowing us to describe the changes of the homogeneous states and the motion of the SFs as they move towards the final stationary state corresponding to a given SF separation.

We have also investigated nonlinear oscillations in symmetry broken ($\theta_F\neq\theta_B$) counterpropagation. We have identified stable limit cycle oscillations in detuning symmetry broken regimes, and observed sudden crisis in which the oscillations become unstable due to a collision with an unstable HSS. Stable oscillatory dynamics coexist with SF solutions for large ranges of parameter values. We have even identified a multi-stability of nonlinear oscillations with SF solutions and the lowest power homogeneous stationary state. This allows for CW and two distinct oscillatory outputs. One can have both fields exhibit slow nonlinear oscillations or a fast switching between homogeneous states present in one field and the other homogeneous.

Micro-resonator systems have undergone much study in recent years. Our predictions have been obtained for realistic parameters with possible experimental verification in a variety of ring resonator setups, from micro-ring to fibre loops. Frequency comb generation has also been demonstrated using two lasers for bichromatic pumping of a micro-ring resonator for the generation of dark bright solitons \cite{PhysRevLett.128.033901}. A modification to this setup to incorporate bidirectional pumping should allow for the generation of counterpropagating SF states. Single input laser setups in the presence of back scattering have predicted and observed Maxwell point front solutions in micro-ring resonators \cite{yu2021continuum,wang2021self}. Back scattering of the pump laser results in a counterpropagating field, allowing a single laser setup to produce plateaus that can be the result of extending our model to these configurations. 

The robust and highly configurable SFs solutions of counterpropagation will be useful in many real world application such as, all optical oscillators, optical computing, time reversal symmetry breaking, signal routing in telecommunication systems.

\vfill

\section*{Acknowledgements}
This research was supported by funding from the Engineering and Physical Sciences Research Council (EPSRC) DTA grant to the University of Strathclyde. PD acknowledges support by the European Union H2020 ERC Starting Grant ``Counterlight" 756966, the Marie Curie Innovative Training Network ``Microcombs'' 812818 and the Max Planck Society.

\bibliographystyle{unsrt} 
\bibliography{biblio.bib}

\appendix
\begin{widetext}

\section{Linear stability of homogeneous stationary states to inhomogeneous perturbation in counterpropagation}\label{app:Linear_stability_homo}
Here we investigate the stability of stationary homogeneous states $F_s,B_s$ to spatial perturbations at zero dispersion ($\beta = 0$). The non-locality of the counterpropagating system means that local perturbations will result in changes to the unperturbed regions, and therefore have an implicit dependence on the entirety of the field. It is necessary to track the evolution of the entire field to determine the susceptibility of the homogeneous stationary states to spatial bifurcation. We do so by considering the field part wise in fast time
\begin{eqnarray}
    F &=& F_1T(\zeta)T(x_F - \zeta) + F_2T(\zeta-x_F)T(L - \zeta)\\
    B &=& B_1T(\zeta)T(x_B - \zeta) + B_2T(\zeta-x_B)T(L - \zeta)
\end{eqnarray}
such that
\begin{eqnarray}
    |F|^2 &=& |F_1|^2T(\zeta)T(x_F - \zeta) + |F_2|^2T(\zeta-x_F)T(L - \zeta)\\
    |B|^2 &=& |B_1|^2T(\zeta)T(x_B - \zeta) + |B_2|^2T(\zeta-x_B)T(L - \zeta)
\end{eqnarray}
where $T(\zeta)$ represent the Heaviside step function which has value 1 for $\zeta\geq 0$, and 0 for $\zeta< 0$, and $x_F,x_B$ are the lengths of fast time occupied by $F_1,B_1$. The part wise fields $F_1$ and $F_2$ ($B_1$ and $B_2$) represent two separate domains of fast time with different spatially homogeneous perturbations of the same HSS, such that the combined perturbation is spatially inhomogeneous. We consider the linear perturbation to the counterpropagating system of the form
\begin{eqnarray}
    F_1 &=& F_s + f_1,\quad F_2 = F_s + f_2\\
    B_1 &=& B_s + b_1,\quad B_2 = B_s + b_2
\end{eqnarray}
The average field powers under this formulation are
\begin{eqnarray}
    \langle |F|^2\rangle &=& \Delta_F|F_1|^2 + (1-\Delta_F)|F_2|^2\\
    \langle |B|^2\rangle &=& \Delta_B|B_1|^2 + (1-\Delta_B)|B_2|^2
\end{eqnarray}
where $\Delta_{F} = x_{F}/L, ~\Delta_{B} = x_{B}/L,$ are the normalised lengths occupied by $F_1,~B_1$. The evolution of the $F_{1}$ and $F_2$ components are not explicitly dependant on each other due to zero dispersion. As such we describe the evolution of the $F$ field as separate ODEs for $F_1,F_2$ (likewise for the $B$ field), hence this system is described by the 4 ODEs
\begin{eqnarray}
    \partial_\tau F_1 &=& S_F - (1 + i\theta_F)F_1 + i(|F_1|^2 + \nu[\Delta_B|B_1|^2 + (1-\Delta_B)|B_2|^2])F_1\\
    \partial_\tau F_2 &=& S_F - (1 + i\theta_F)F_2 + i(|F_2|^2 + \nu[\Delta_B|B_1|^2 + (1-\Delta_B)|B_2|^2])F_2\\
    \partial_\tau B_1 &=& S_B - (1 + i\theta_B)B_1 + i(|B_1|^2 + \nu[\Delta_F|F_1|^2 + (1-\Delta_F)|F_2|^2])B_1\\
    \partial_\tau B_2 &=& S_B - (1 + i\theta_B)B_2 + i(|B_2|^2 + \nu[\Delta_F|F_1|^2 + (1-\Delta_F)|F_2|^2])B_2
\end{eqnarray}
Without loss of generality, we adjust the phase of $F,B$ such that $F_s,B_s$ are real. We have that the real and imaginary components of the perturbation evolve as
\begin{align}
\frac{d}{d\tau}
\begin{pmatrix}
f_{1,r}\\
f_{1,i}\\
f_{2,r}\\
f_{2,i}\\
b_{1,r}\\
b_{1,i}\\
b_{2,r}\\
b_{2,i}\\
\end{pmatrix}
=
\begin{pmatrix}
-1 & A_1 & 0 & 0 & 0 & 0 & 0 & 0\\
-B_1 & -1 & 0 & 0 & -\Delta_BC & 0 & -(1-\Delta_B)C & 0\\
0 & 0 & -1 & A_1 & 0 & 0 & 0 & 0\\
0 & 0 & -B_1 & -1 & -\Delta_BC & 0 & -(1-\Delta_B)C & 0\\
0 & 0 & 0 & 0 & -1 & A_2 & 0 & 0\\
-\Delta_FC & 0 & -(1-\Delta_F)C & 0 & -B_2 & -1 & 0 & 0 \\
0 & 0 & 0 & 0 & 0 & 0 & -1 & A_2\\
-\Delta_FC & 0 & -(1-\Delta_F)C & 0 & 0 & 0 & -B_2 & -1\\
\end{pmatrix}
\begin{pmatrix}
f_{1,r}\\
f_{1,i}\\
f_{2,r}\\
f_{2,i}\\
b_{1,r}\\
b_{1,i}\\
b_{2,r}\\
b_{2,i}\\
\end{pmatrix}
\end{align}
where $A_1 = F_s^2 + \nu B_s^2 - \theta_F$, $A_2 = B_s^2 + \nu F_s^2 - \theta_B$, $B_1 = 3F_s^2 + \nu B_s^2 - \theta_F$, $B_2 = 3B_{s}^2 + \nu F_s^2 - \theta_B$, and $C = 2\nu F_sB_s$. This results in the known eigenvalues of homogeneous perturbation of the homogeneous stationary states \cite{woodley2018universal}
\begin{eqnarray}
    \lambda &=& -1 \pm \frac{\sqrt{-A_1B_1-A_2B_2\pm S}}{\sqrt{2}}\label{eq:homo_eigenvalue}\\
    S &=& \sqrt{(A_1B_1 - A_2B_2)^2 + 4A_1A_2C^2}
\end{eqnarray}
with addition eigenvalues indicative of instability of either the $F$ field ($\lambda_F$) or the $B$ field ($\lambda_B$) due to spatially inhomogeneous perturbations
\begin{eqnarray}
    \lambda_F &=& -1 \pm \sqrt{-A_1B_1}\label{eq:app_spatialF}\\
    \lambda_B &=& -1 \pm \sqrt{-A_2B_2}\label{eq:app_spatialB}
\end{eqnarray}
These 4 eigenvalues are a consequence of the nonlocal coupling and are not present in local coupling regime of copropagating fields. They are identical to those seen in 2 single LLEs with parameter values $P_F,\Tilde{\theta}_F$ \& $P_B,\Tilde{\theta}_B$. We note that $F_1,F_2$ do not need to be continuous regions of fast time. They represent the total proportion of the field perturbed below or above the stationary solution and as such the above eigenvalues are appropriate for a random spatial perturbation (which would have width $\Delta_{F}\approx 0.5$). Likewise for the $B$ field.

In the regime of local coupling, the two copropagating fields are coupled by Kerr cross phase modulation. As such a local spatial perturbation of one of the fields will only effect the corresponding spatial region of the other field. If we introduce a step function perturbation to the homogeneous stationary states $F_s,B_s$ with size $\Delta$ of the form $F = F_s + fT(\zeta)T(\Delta - \zeta/L),~ B = B_s + bT(\zeta)T(\Delta - \zeta/L)$, the perturbations $f,b$ will evolve identically to a homogeneous perturbation of the entire field. This results in the eigenvalues given by Eq. (\ref{eq:homo_eigenvalue}) of homogeneous perturbation of the homogeneous stationary states \cite{woodley2018universal}. Non-locality in the counterpropagating system introduces an implicit dependence on the power of the entire field. This allows the system to access inhomogeneous states of the single LLE, and introduces 4 additional eigenvalues indicative of spatial instability.

\begin{figure}[]
    \centering
    {\includegraphics[width=0.5\textwidth]{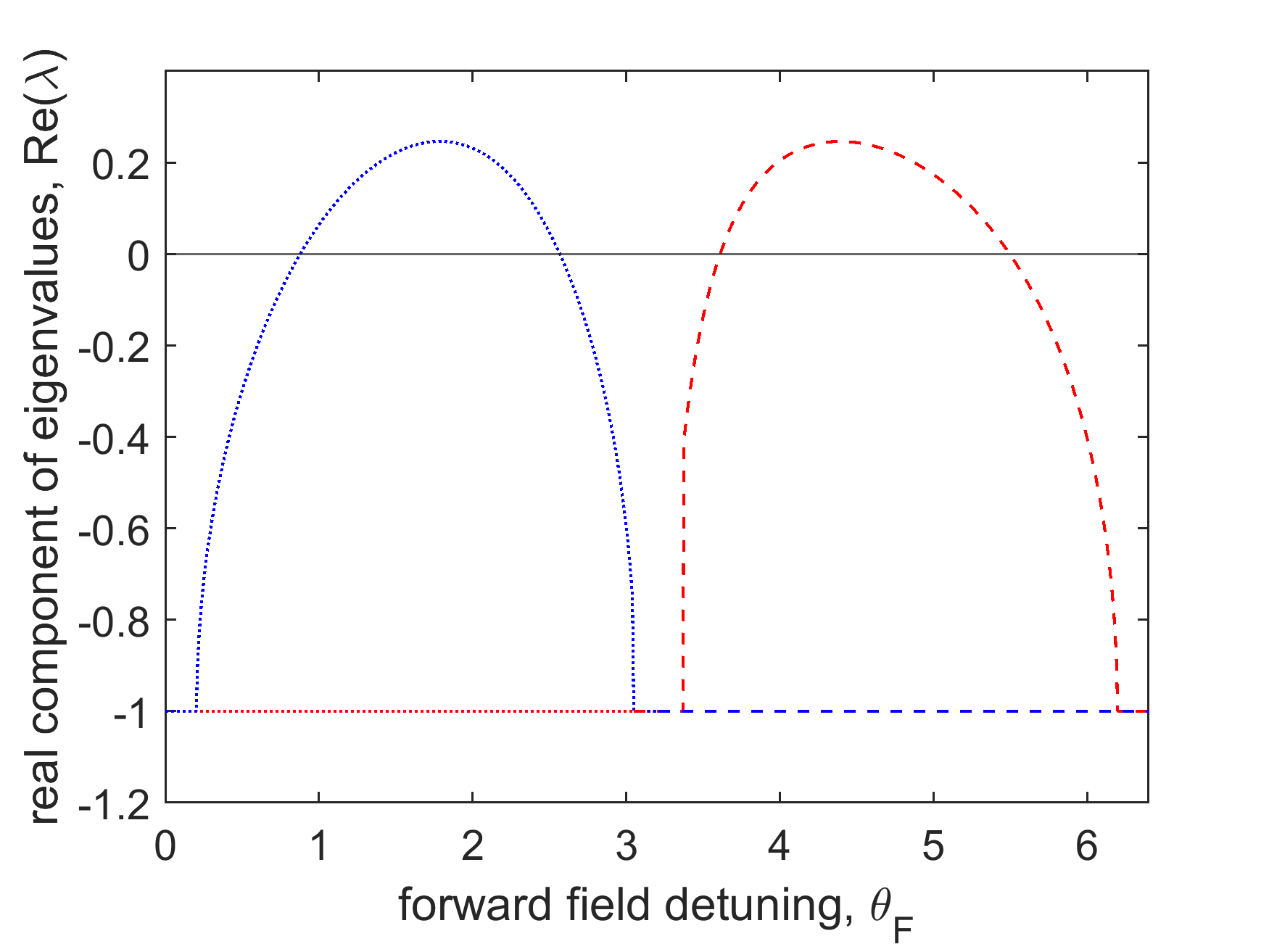}}
    \caption{Real component of eigenvalues for changing $\theta_F$ with parameter values $P_F=P_B=2.1609,\theta_B=3.2$. The corresponding HSS are plotted in Fig. \ref{fig:HSS}. For $\theta_F>\theta_B$ (dashed lines) we consider the highest power branch of HSS. For $\theta_F<\theta_B$ (dotted lines) we consider the sole HSS. The real components of the `+' solutions of Eqs. (\ref{eq:app_spatialF})\&(\ref{eq:app_spatialB}) indicate instability. Plotted above are the `+' solutions of $\lambda_F$ (red) and $\lambda_B$ (blue).}
    \label{}
\end{figure}

\section{Linear stability of inhomogeneous front stationary states in counterpropagating fields}\label{app:linear_stability_inhomo}
In numerical simulations, we observe that stationary SFs form in only one field at any a given time, with the other field remaining homogeneous. Using a similar framework as in Appendix \ref{app:Linear_stability_homo}, we can simply do the analysis by considering a homogeneous $F$ field with an inhomogeneous $B$ field. We describe the $B$ field as the part wise function in terms of the higher and lower power homogeneous state $B^+,B^-$ connected by the SFs, and the $F$ field as a single homogeneous function. At zero dispersion we have
\begin{eqnarray}
    B &=& B^-T(\zeta)T(\Delta_B - \zeta/L) + B^+T(\zeta/L-\Delta_B)T(1 - \zeta/L)
\end{eqnarray}
where $\Delta$ is the normalised front separation. Therefore
\begin{eqnarray}
    |F|^2 &=& |F|^2\\
    |B|^2 &=& |B^-|^2T(\zeta)T(\Delta_B - \zeta/L) + |B^+|^2T(\zeta/L-\Delta_B)T(1 - \zeta/L)
\end{eqnarray}
and the average field power is
\begin{eqnarray}
    \langle|F|^2\rangle &=& |F|^2\\
    \langle|B|^2\rangle &=& \Delta_B|B^-|^2 + (1-\Delta_B)|B^+|^2
\end{eqnarray}
As the $B$ field is part wise and the $F$ field is homogeneous, the evolution of the $F,B$ fields is described by the 3 ODEs
\begin{eqnarray}
    \partial_\tau F &=& S_F - (1 + i\theta_F)F + i(|F|^2 + \nu[\Delta_B|B^-|^2 + (1-\Delta_B)|B^+|^2])F\\
    \partial_\tau B^+ &=& S_B - (1 + i\theta_B)B^+ + i(|B^+|^2 + \nu|F|^2)B^+\\
    \partial_\tau B^- &=& S_B - (1 + i\theta_B)B^- + i(|B^-|^2 + \nu|F|^2)B^-
\end{eqnarray}
We introduce a linear perturbation to the system that is spatially inhomogeneous in the $B$ field and homogeneous in the $F$ field,
\begin{eqnarray}
    F &=& F_s + f\\
    B^+ &=& B_s^+ + b^+,\quad B^- = B_s^- + b^-
\end{eqnarray}
where $F_s$ is the stationary homogeneous solution of the $F$ field and \& $B^+_s,~B^-_s$ are the two homogeneous stationary states connected by the SFs. Without loss of generality, we adjust the phase of $F,B$ such that $F_s,B_s$ are real. We have that the real and imaginary components of the perturbations evolve as
\begin{align}
\frac{d}{d\tau}
\begin{pmatrix}
f_{r}\\
f_{i}\\
b_{r}^+\\
b_{i}^+\\
b_{r}^-\\
b_{i}^-
\end{pmatrix}
=
\begin{pmatrix}
-1 & A & 0 & 0 & 0 & 0\\
-B & -1 & -(1-\Delta_B)C_1 & 0 & -\Delta_BC_2 & 0\\
0 & 0 & -1 & A_1 & 0 & 0\\
-C_1 & 0 & -B_1 & -1 & 0 & 0\\
0 & 0 & 0 & 0 & -1 & A_2\\
-C_2 & 0 & 0 & 0 & -B_2 & -1
\end{pmatrix}
\begin{pmatrix}
f_{r}\\
f_{i}\\
b_{r}^+\\
b_{i}^+\\
b_{r}^-\\
b_{i}^-
\end{pmatrix}
\end{align}
where $A = F_s^2 + \nu\langle |B_s|^2\rangle - \theta_F,~ B = 3F_s^2 + \nu\langle |B_s|^2\rangle - \theta_F,~ A_1 = (B_s^+)^2 + \nu F_s^2 - \theta_B,~ B_1 = 3(B_s^+)^2 + \nu F_s^2 - \theta_B,~ A_2 = (B_s^-)^2 + \nu F_s^2 - \theta_B,~ B_2 = 3(B_s^-)^2 + \nu F_s^2 - \theta_B,~ C_1 = 2\nu F_sB_s^+,~ C_2 = 2\nu F_sB_s^-$. This results in the characteristic polynomial
\begin{equation}
  \begin{split}
    0 &= [(\lambda+1)^2+A_2B_2]\Big\{[(\lambda+1)^2+AB][(\lambda+1)^2+A_1B_1] - 2\Delta_BAA_1C_1^2\Big\}\\
    &+ [(\lambda+1)^2+A_1B_1]\Big\{[(\lambda+1)^2+AB][(\lambda+1)^2+A_2B_2]-2(1-\Delta_B)AA_2C_2^2\Big\}
  \end{split}\label{eq:app_TFSSstability}
\end{equation}
which is composed of the product of terms indicative of spatial instability
\begin{eqnarray}
    \Lambda_n^\pm &=& \lambda + 1 \pm \sqrt{-A_nB_n}
\end{eqnarray}
and eigenvalues indicative of temporal instability (in the curly brackets)
\begin{eqnarray}
    L_n^{(\pm,\pm)} &=& \lambda + 1 \pm \frac{\sqrt{-AB-A_nB_n\pm S_n}}{\sqrt{2}},\quad S_n = \sqrt{(AB - A_nB_n)^2 + (-1)^n(1 - n - \Delta_B) 8AA_nC_n^2}
\end{eqnarray}
This expression has similar form the the characteristic polynomial of HSS seen in appendix \ref{app:Linear_stability_homo} and will become identical when $\Delta_B = 0,1$. In simulation, we observe that the SF solutions are susceptible to damped oscillations under perturbation. These oscillations grow in the range $3.5<\theta_F<6.3$ and the SF solutions are unstable.

\begin{figure}[!h]
    \centering
    {\includegraphics[width=0.5\textwidth]{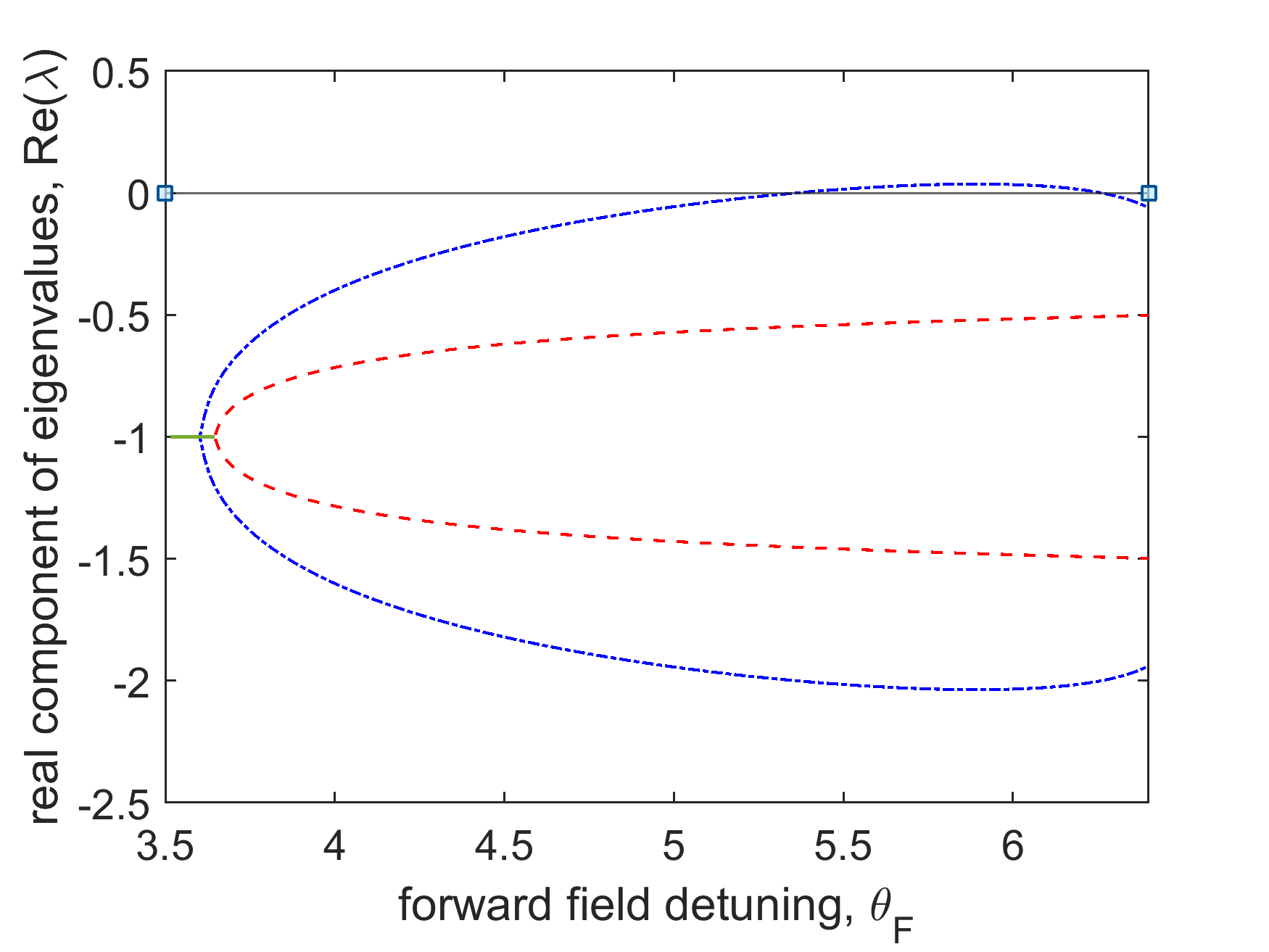}}
    \caption{Real component of eigenvalues of the zero dispersion SF solutions for changing $\theta_F$ with parameter values $P_F=P_B=2.1609,\theta_B=3.2$. The six eigenvalues are calculated numerically (six roots of Eq. (\ref{eq:app_TFSSstability})), where each branch of the blue dot dashed line represent the real part of a complex conjugate pair of solutions, the red dashed lines are real solutions.}
\end{figure}

If we instead consider a spatially inhomogeneous perturbation to the homogeneous plateau of the SF solution present in the backward field, that does not change the average power of the field $\langle|B_s + b(\zeta)|^2\rangle = \langle|B_s|^2\rangle$, then the resulting eigenvalues are
\begin{eqnarray}
    \lambda &=& - 1 \pm \sqrt{-AB},\quad\quad\text{forward field}\\
    \lambda_n &=& - 1 \pm \sqrt{-A_nB_n},\quad\text{backward field}
\end{eqnarray}
These eigenvalues are indicative of the spatial stability of the two HS connected to the SFs. This suggests that temporal instability of the stationary states of counterpropagating fields is observed when the integrated powers of the fields are perturbed. Otherwise the fields exhibit the stability of an LLE with effective detuning as defined in section \ref{sec:TFSS_distance}. In particular, the eigenvalues of the SF solution at stationary separation as calculated in section \ref{sec:TFSS_distance} are those of a single LLE at Maxwell point
\begin{eqnarray}
    \lambda^+ &=& -1 \pm \sqrt{(\Theta_{MP}-Y_B^+)(3Y_B^+-\Theta_{MP})}\\
    \lambda^- &=& -1 \pm \sqrt{(\Theta_{MP}-Y_B^-)(3Y_B^--\Theta_{MP})}
\end{eqnarray}
This is expected due to to the one to one correspondence of the counterpropagating SF solution to the stationary states of the LLE. We note that these eigenvalues are independent of the detuning values. As such the solutions map to the identical Maxwell point LLE when changing $\theta_F$ which exhibits a multi-stability of SF states. The eigenvalues of the forward field are 
\begin{eqnarray}
    \lambda &=& -1 \pm \sqrt{(\Tilde{\theta}_F-Y_F)(3Y_F-\Tilde{\theta}_F)}
\end{eqnarray}
where $\Tilde{\theta}_F = \theta_F - \nu\langle|B_s|^2\rangle$ is the effective detuning which is dependent on the detuning values (or more specifically the front separation $\Delta$).

\end{widetext}

\end{document}